\documentclass[aps,prd,unsortedaddress,superscriptaddress,showpacs,nofootinbib,11pt]{revtex4-2}

\pdfoutput=1

\usepackage{graphicx,color}
\usepackage{relsize}
\usepackage{slashed}
\usepackage[normalem]{ulem}
\usepackage{tabu}
\usepackage{rotating}
\usepackage{bigstrut}
\usepackage{makecell}
\usepackage[dvipsnames]{xcolor}

\usepackage{amsmath}
\usepackage{amssymb,bm}
\usepackage{amsthm}
\usepackage{mathrsfs}
\usepackage{fancyhdr}
\usepackage{array}
\usepackage[all]{xy}
\usepackage{eufrak}
\usepackage{euscript}
\usepackage{enumerate}
\usepackage{slashed}
\usepackage{hyperref}
\usepackage{subfigure} 
\usepackage{mathtools}

\usepackage{soul}

\hypersetup{pdftex,colorlinks=true,linkcolor=blue,citecolor=blue,menucolor=black,urlcolor=blue,filecolor=blue}

\newcommand{\itp}{\affiliation{CAS Key Laboratory of Theoretical Physics, Institute of Theoretical Physics,\\
Chinese Academy of Sciences, Beijing 100190, China}}
\newcommand{\ucas}{\affiliation{School of Physical Sciences, University of Chinese Academy of Sciences,\\ Beijing 100049, China}}
\newcommand{\imp}{\affiliation{Institute of Modern Physics, Chinese Academy of Sciences, Lanzhou 730000, China}}

\newcommand{\peng}{\affiliation{Peng Huanwu Collaborative Center for Research and Education, Beihang University, Beijing 100191, China}}

\begin{document}

\title{On the $\eta_1(1855)$, $\pi_1(1400)$ and $\pi_1(1600)$ as dynamically generated states and their SU(3) partners\footnote{https://doi.org/10.3390/universe9020109}}

\author{Mao-Jun Yan}
\email{yanmaojun@itp.ac.cn}
\itp

\author{Jorgivan M. Dias}
\email{jorgivan.mdias@itp.ac.cn}
\itp

\author{ Adolfo Guevara}
\email{aguevara@itp.ac.cn}
\itp

\author{Feng-Kun Guo}
\email{fkguo@itp.ac.cn}
\itp\ucas \peng

\author{Bing-Song Zou}
\email{zoubs@itp.ac.cn}
\itp \ucas \imp 



\begin{abstract}
In this work, we interpret the newly observed $\eta_1(1855)$ 
resonance with exotic $J^{PC}=1^{-+}$ 
quantum numbers in the $I=0$ sector, reported by the BESIII Collaboration, as 
a dynamically generated state from the interaction between the lightest pseudoscalar mesons and
{axial-}vector mesons. The interaction is 
derived from the lowest order chiral Lagrangian from 
which the Weinberg-Tomozawa term is obtained, describing 
the transition amplitudes among the relevant channels, 
which are then unitarized using the Bethe-Salpeter 
equation, according to the chiral unitary approach. 
We evaluate the $\eta_1(1855)$ 
decays into the $\eta\eta^{\prime}$ and $K\bar{K}^*\pi$ 
channels and find that the latter has a larger branching fraction. 
We also investigate its SU(3) partners, 
and according to our findings, the $\pi_1(1400)$ and $\pi_1(1600)$ 
structures may correspond to dynamically generated states, with 
the former one coupled mostly to the $b_1\pi$ component and the latter one 
coupled to the $K_1(1270)\bar{K}$ channel. In particular, our result for 
the ratio $\Gamma(\pi_1(1600)\to f_1(1285)\pi)/ \Gamma(\pi_1(1600)\to \eta^{\prime}\pi)$ 
is consistent with the measured value, which supports our 
interpretation for the higher $\pi_1$ state. We also report two poles with a mass about 1.7~GeV
in the $I=1/2$ sector, which may be responsible for the $K^*(1680)$. We suggest searching for two additional  $\eta_1$ exotic mesons with masses around 1.4 and 1.7~GeV. In particular, the predicted $\eta_1(1700)$ is expected to have a width around 0.1~GeV and can decay easily into $K\bar K\pi\pi$. 
\end{abstract}

\maketitle

\newpage

\section{Introduction}
\label{sec:intro}

Over the last two decades, the experimental 
observation of many new hadronic states is challenging 
our current understanding of hadrons as conventional mesons and 
baryons with valence contents of quark-antiquark and three quarks, respectively, since most of them do not fit in the well-known 
quark model. 
This difficulty 
brought back a long-standing discussion on the exotic 
hadronic structures, {\it i.e.}, multiquark 
configurations that might have quantum numbers beyond 
those assigned to the conventional mesons and baryons~\cite{Chodos:1974je,Jaffe:1976ig}.

Exotic quark configurations 
such as tetraquarks \cite{Maiani:2004vq,Esposito:2016noz}, 
hadron-hadron molecules \cite{Guo:2017jvc}, glueballs, and 
hybrids \cite{Luo:2005zg,Meyer:2015eta}, 
among others, have been suggested to describe suitably 
most of the properties of these new states, such 
as the $J^{PC}$ quantum numbers, mass, and decay width, 
especially for those lying in the charmonium and 
bottomonium spectra.

On the other hand, distinguishing the exotic states 
from the conventional hadrons is a more complicated task 
in the light quark sector. Many states have their masses 
close to each other, and the possibility of mixing 
brings additional difficulty to the problem. 
The situation improves as the quantum numbers do not 
fall into those allowed by the conventional quark model. 
It seems to be the case of the newly discovered state, dubbed 
$\eta_1(1855)$, by the BESIII Collaboration~\cite{BESIII:2022riz,BESIII:2022zel},
observed in the invariant mass distribution of the $\eta\,\eta^{\prime}$ 
meson pair in the $J/\psi\to \gamma\, \eta\,\eta^{\prime}$ 
decay channel with a significance of $19\sigma$. Its mass and 
width reported by BESIII are $1855\pm 9^{+6}_{-1}$~MeV and 
$188\pm 18^{+3}_{-8}$~MeV, respectively, with likely 
$J^{PC}=1^{-+}$ quantum numbers, which cannot be formed by a pair of quark and antiquark. The $\eta_1(1855)$ is not 
the only state experimentally found with that set of quantum 
numbers. As of today, three other hadronic structures, 
called $\pi_1(1400)$, $\pi_1(1600)$ and $\pi_1(2015)$, 
with $J^{PC}=1^{-+}$, were observed by several 
collaborations~\cite{Meyer:2015eta,ParticleDataGroup:2022pth}.

From the theoretical point of view, the hybrid model has been used 
to investigate these exotic meson states, in particular the $1^{- +}$ ones. Lattice quantum chromodynamics (QCD) calculations 
have pointed out hybrid supermultiplets with exotic $J^{PC}$ 
quantum numbers, including the $1^{-+}$ one~\cite{Lacock:1996vy,Lacock:1996ny,MILC:1997usn,McNeile:2006bz,Dudek:2010wm,Dudek:2011bn}.
In this picture, however, the mass of the lightest $1^{-+}$ state and decay modes  are inconsistent 
with the corresponding experimental results,  while the $\pi_1(1600)$ 
and $\pi_1(2015)$ structures can fit into the nonets predicted 
by lattice QCD~\cite{Meyer:2015eta}.

The newly observed $\eta_1(1855)$ state has also been the focus of some 
studies. In particular, the authors in Ref.~\cite{Qiu:2022ktc} proposed 
two hybrid nonet schemes in which the $\eta_1(1855)$ resonance can 
be either the lower or higher mass state with isospin $I=0$. 
In Ref.~\cite{Shastry:2022mhk}, an effective Lagrangian 
respecting flavor, parity, and charge conjugation symmetries is used to study the 
hybrid nonet decays into two-body meson states. The 
authors have fixed the couplings to those two-body meson 
states by performing a combined fit to the experimental and 
lattice results available. As a result, the decay width value 
estimated for the isoscalar member of the hybrid nonet agrees 
with the one observed for $\eta_1(1855)$ state. Also addressing 
the same picture, Ref.~\cite{Chen:2022qpd} applied 
the approach of QCD sum rules to describe the $\eta_1(1855)$ mass. 
By contrast, within the same approach, the $\eta_1(1855)$ 
resonance is described as a tetraquark state in 
Ref.~\cite{Wan:2022xkx}.

The $\eta_1(1855)$ resonance also 
supports a meson-meson molecule interpretation due to its 
proximity to the $K\bar{K}_1(1400)$ threshold, as put 
forward by Refs.~\cite{Dong:2022cuw,Yang:2022lwq}. 
In particular, the authors in Ref.~\cite{Dong:2022cuw} 
have investigated the $K\bar{K}_1(1400)$ interaction 
through the one-boson exchange model. According to their 
findings, the $K\bar{K}_1(1400)$ system binds for cutoff 
values above $2$ GeV with a monopole form factor. In addition, the comparison 
between their result for the branching fraction 
$\mathcal{B}(\eta_1 \to \eta\,\eta^{\prime})$ to the 
experimental one led them to conclude that the $K\bar{K}_1(1400)$ 
molecule can explain the $\eta_1(1855)$ structure. 

An important point to be addressed is the meson-meson 
interaction around the $K_1(1400) \bar{K}$ threshold for the 
$J^{PC}=1^{-+}$ quantum numbers. In this sector, many meson-meson pairs 
may contribute to that interaction, so a coupled-channel 
treatment seems appropriate to take these contributions 
into account. In particular, hadron-hadron interactions 
in coupled channels have been studied in many works to 
describe the properties of the new hadronic systems experimentally 
observed. In those cases, these hadronic structures are 
called dynamically generated states.

Following this approach, in this work, we aim to explore the $\eta_1(1855)$, 
$\pi_1(1400)$, and $\pi_1(1600)$ hadronic systems as dynamically 
generated states from pseudoscalar-axial vector meson 
interactions in coupled channels. Specifically, the low-energy interactions 
are given by the Weinberg-Tomozawa (WT) term 
from chiral Lagrangians at the leading order of the chiral expansion by treating the axial vector mesons as matter fields and the pseudoscalar mesons as the pseudo-Nambu-Goldstone bosons of the spontaneous breaking of chiral symmetry. Such Lagrangians have been  used to study many hadron structures stemming from 
meson-meson and meson-baryon  interactions in 
coupled channels in light and heavy sectors, see, e.g., Refs.~\cite{Oller:2000ma,Hofmann:2003je,Roca:2005nm,Guo:2006fu,Dias:2018qhp}. 
In our case, the amplitudes obtained from the WT 
term are unitarized via the Bethe-Salpeter equation 
from which bound states/resonances manifest as poles 
in the physical/unphysical Riemann sheets of the scattering 
matrices. 
The existence of a whole family of kaonic bound states has been pointed out in Ref.~\cite{Guo:2011dd} based on unitarizing the WT term for the scattering of the kaon off isospin-1/2 matter fields taking heavy mesons and doubly-charmed baryons as examples.
As we shall show in this work, the newly observed
$\eta_1(1855)$ structure may correspond to a dynamically 
generated state from the pseudoscalar-axial vector interaction 
in the isospin $I=0$ sector coupling strongly to the $K_1(1400)\bar{K}$ 
channel. Moreover, the $\pi_1(1400)$ and 
$\pi_1(1600)$, may be assigned as the $\eta_1(1855)$ SU(3) 
partners which are also dynamically generated from the pseudoscalar-axial 
vector meson interactions in the $I=1$ sector. The former resonance couples mainly to the $b_1\pi$ channel, and the latter has the $K_1(1270)\bar{K}$ as its main coupled channel.

In addition, we have also found two poles around 1.7~GeV in the $I=1/2$ sector. 
These poles are particularly interesting as they could be the origin of the $K^*(1680)$ structure observed experimentally~\cite{ParticleDataGroup:2022pth}, which is the main component of the $1^-$ contribution to the $\phi K$ mass distribution in the $B\to J/\psi\phi K$  decays recently measured by LHCb~\cite{LHCb:2021uow}.

 
This paper is organized as follows. In Section \ref{sec:formalism}, 
we discuss the relevant channels contributing to the pseudoscalar-axial 
vector meson interactions and the use of the chiral unitary approach (ChUA) for the evaluation of the transition amplitudes among those 
channels. 
In Sections \ref{sec:eta} and \ref{sec:pis}, we investigate the 
dynamical generation of poles stemming from those interactions in 
the $I=0$ and $I=1$ sectors and discuss their possible decay channels. 
Finally, in Section \ref{sec:ihalf}, we also explore the dynamical 
generation of poles for $I=1/2$ and their connection to the vector $K^*(1680)$ structure observed experimentally. 
Section \ref{sec:conc} gives a summary. 

\section{Coupled channel scattering in chiral unitary
approach}
\label{sec:formalism}

We investigate the 
interactions between axial and pseudoscalar mesons in 
coupled channels in the $1300 \sim 2000$~MeV energy range. 
First, we need to determine the space of states contributing 
to the interaction in this energy range.

In Tables \ref{tab1}, \ref{tab2}, \ref{tab3},  and \ref{tab4}, 
we list all the relevant channels for the problem under 
consideration along with their corresponding mass thresholds. 
The channels are organized from the lower to higher mass values 
and by the isospin, $0$, $1$ and $1/2$, respectively. 

\begin{table}[h!]
\caption{$J^{PC}=1^{-+}$ meson-meson channels with $I=0$. The threshold masses are in the units of MeV.}
\centering
\begin{tabular}{l | c c c c c}
\hline\hline
{Channel} ~& ~$a_1\pi$~ & ~$K_1(1270)\bar{K}$~ & ~$f_1(1285)\eta$~ & ~$K_1(1400)\bar{K}$ ~&~ $f_1(1420)\eta$\\
\hline
{Threshold} ~& $1368$ & $1748$ & $1829$ & $1898$ & $1973$\\
\hline\hline
\end{tabular}
\label{tab1}
\end{table}

\begin{table}[h!]
\caption{$J^{PC}=1^{-+}$ meson-meson channels with $I=1$.  The threshold masses are in the units of MeV.}
\centering
\begin{tabular}{l | c c c c c c}
\hline\hline
{Channel} ~& ~$b_1\pi$~ & ~$f_1(1285)\pi$~ & ~$f_1(1420)\pi$ ~&~$K_1(1270)\bar{K}$ ~& ~$a_1\eta$ ~&~ $K_1(1400)\bar{K}$\\
\hline
{Threshold} ~& $1367$ & $1419$ & $1564$ & $1748$ & $1777$ & $1895$\\
\hline\hline
\end{tabular}
\label{tab2}
\end{table} 

\begin{table}[h!]
\caption{$J^{P}=1^{-}$ meson-meson channels with $I=1/2$. The threshold masses are in the units of MeV. Here the flavor-neutral axial vector mesons have $J^{PC}=1^{++}$.}
\centering
\begin{tabular}{l | c c c c c}
\hline\hline
{Channel} ~& ~$a_1 K$~ & ~$f_1(1285)K$~ & ~$K_1(1270) \eta$~ & ~$f_1(1420) K$ ~&~ $K_1(1400)\eta$\\
\hline
{Threshold} ~& $1725$ & $1777$ & $1800$ & $1921$ & $1947$\\
\hline\hline
\end{tabular}
\label{tab3}
\end{table}

\begin{table}[h!]
\caption{$J^{P}=1^{-}$ meson-meson channels with $I=1/2$. The threshold masses are in the units of MeV. Here the flavor-neutral axial vector mesons have $J^{PC}=1^{+-}$.}
\centering
\begin{tabular}{l | c c c c c}
\hline\hline
{Channel} ~& ~$h_1(1170) K$~ & ~$b_1 K$~ & ~$K_1(1270) \eta$~ & ~$h_1(1415) K$ ~&~ $K_1(1400)\eta$\\
\hline
{Threshold} ~& $1661$ & $1725$ & $1800$ & $1911$ & $1947$\\
\hline\hline
\end{tabular}
\label{tab4}
\end{table}

In what follows, we shall discuss the relevant scattering 
amplitudes among all those channels above for each isospin 
sector. These transitions can be written in the form of 
the WT term which then is unitarized. 
Notice that the channels displayed in Tables \ref{tab3} and \ref{tab4}, in principle, should be grouped in the same space of states since they share identical isospin and $J^P$ quantum numbers. However, the relevant transitions among them arise only at the next-to-leading order in the chiral expansion; see the discussion around Eq.~\eqref{eq:a1b1} below. 
Thus, such transitions are of higher order than that of the WT term and will be neglected here.

\subsection{The Weinberg-Tomozawa term}
\label{subsec:amp}

In order to study the interactions among all the 
channels listed in the previous tables, 
we have to evaluate the interactions between the pseudoscalar 
and axial-vector mesons. The latter are organized in 
two SU(3) {octets} according to their $J^{PC}$ quantum numbers. 
\begin{align}
\label{amatrix}
&A_{1}=\left(\begin{array}{ccc}
\frac{a_{1}^{0}}{\sqrt{2}}+\frac{f_{1}^{8}}{\sqrt{6}} & a_{1}^{+} & K_{1 A}^{+} \\
a_{1}^{-} & -\frac{a_{1}^{0}}{\sqrt{2}}+\frac{f_{1}^{8}}{\sqrt{6}} & K_{1 A}^{0} \\
K_{1 A}^{-} & \bar{K}_{1 A}^{0} & -\frac{2 f_{1}^{8}}{\sqrt{6}}
\end{array}\right) 
\end{align}
is the octet  of resonances of axial-vector 
states with $J^{PC}=1^{++}$ for the flavor-neutral mesons, and
\begin{align}
\label{bmatrix}
B_{1}=\left(\begin{array}{ccc}
\frac{b_{1}^{0}}{\sqrt{2}}+\frac{h_{1}^8}{\sqrt{6}} & b_{1}^{+} & K_{1 B}^{+} \\
b_{1}^{-} & -\frac{b_{1}^{0}}{\sqrt{2}}+\frac{h_{1}^8}{\sqrt{6}} & K_{1 B}^{0} \\
K_{1 B}^{-} & K_{1 B}^{0} & -\frac{2}{\sqrt{6}} h_{1}^8
\end{array}\right) 
\end{align} 
describes the octet of axial-vector resonances with $J^{PC}=1^{+-}$. 
The singlet and $I=0$ octet flavor eigenstates are not mass eigenstates;  
that is, the pairs of $f_1(1420)$, 
$h_1(1415)$ (also known as $h_1(1380)$) and $f_1(1285)$, 
$h_1(1170)$ mesons are 
mixtures of the singlet $(^{1})$ and octet $(^8)$ mesons such that 
\begin{align}
\left(\begin{array}{c}
\left|f_{1}(1285)\right\rangle \\
\left|f_{1}(1420)\right\rangle
\end{array}\right)=\left(\begin{array}{cc}
\cos \theta_{^3P_1} & \sin \theta_{^3P_1} \\
- \sin \theta_{^3P_1} & \cos\theta_{^3P_1}
\end{array}\right)\left(\begin{array}{l}
\left|f^1_1\right\rangle \\
\left|f^8_1\right\rangle
\end{array}\right) ,
\label{f1}
\end{align}
and 
\begin{align}
\left(\begin{array}{c}
\left|h_{1}(1170)\right\rangle \\
\left|h_{1}(1415)\right\rangle
\end{array}\right)=\left(\begin{array}{cc}
\cos \theta_{^1P_1} & \sin \theta_{^1P_1} \\
- \sin \theta_{^1P_1} & \cos\theta_{^1P_1}
\end{array}\right)\left(\begin{array}{l}
\left|h^1_1\right\rangle \\
\left|h^8_1\right\rangle
\end{array}\right) .
\label{h1}
\end{align}
Furthermore, the $K_{1A}$ and $K_{1B}$ members of the 
multiplets in Eqs.~\eqref{amatrix} and \eqref{bmatrix} 
are the strange partners of the $a_1(1260)$ and $b_1(1235)$, 
and their mixture contributes to the physical $K_1(1270)$ and 
$K_1(1400)$ mesons, that is
\begin{align}
\left(\begin{array}{c}
\left|K_{1}(1270)\right\rangle \\
\left|K_{1}(1400)\right\rangle
\end{array}\right)=\left(\begin{array}{cc}
\sin \theta_{K_1} & \cos \theta_{K_1} \\
\cos \theta_{K_1} & -\sin \theta_{K_1}
\end{array}\right)\left(\begin{array}{l}
\left|K_{1 A}\right\rangle \\
\left|K_{1 B}\right\rangle
\end{array}\right) .
\label{k1ab}
\end{align}
The corresponding values for the mixing angles in 
Eqs.~\eqref{f1}, \eqref{h1}, and \eqref{k1ab} are listed 
in Table~\ref{angles}, where they are grouped into two 
sets, denoted by A and B. Although set B is preferred in Ref.~\cite{Cheng:2011pb}, we will use both sets to have an estimate of the uncertainties caused by such an angle.

\begin{table}[tb]
\caption{Two sets of values of the axial-vector meson mixing angles taken from Ref.~\cite{Cheng:2011pb}. Set B is preferred in Ref.~\cite{Cheng:2011pb}. The $\eta$-$\eta'$ mixing angle  $\theta_P$ is taken from Ref.~\cite{Amsler:1995td}. For more discussions about these mixing angles, we refer to the review of {\it Quark Model} in the {\it Review of Particle Physics}~\cite{ParticleDataGroup:2022pth}. }
\centering
\begin{tabular}{c | c c c c}
\hline\hline
{Angles} ~& ~$\theta_{K_1}$~ & ~$\theta_{^3P_1}$~ & ~$\theta_{^1P_1}$~ & ~$\theta_P$\\
\hline
{Set A} ~& $57^\circ$ & $52^\circ$ & $-17.5^\circ$ & $-17^\circ$\\
\hline 
{Set B} ~& $34^\circ$ & $23.1^\circ$ & $28.0^\circ$ & $-17^\circ$\\
\hline\hline
\end{tabular}
\label{angles}
\end{table}

In order to determine the WT term we start with 
the Lagrangian (see, e.g., Ref.~\cite{Meissner:1987ge})
\begin{equation}
\mathcal{L}_{0}=-\frac{1}{4} \left\langle V_{\mu \nu} 
V^{\mu \nu}-2 M_{V}^{2} V_{\mu} V^{\mu}\right\rangle ,
\end{equation}
where $\langle,\rangle$ takes trace in the SU(3) flavor space,
\begin{equation}
V_{\mu \nu}=\mathcal{D}_{\mu} V_{\nu}-\mathcal{D}_{\nu} V_{\mu}\, ,
\end{equation}
while $\mathcal{D}_{\mu}$ is the chirally covariant derivative, which when acting on SU(3) octet matter fields reads as
\begin{equation}
\mathcal{D}_{\mu}=\partial_{\mu}+\left[\Gamma_{\mu},\,\right] ,
\end{equation}
with $[\,\,,\,\,]$ the usual commutator. In addition, 
$\Gamma_{\mu}$ stands for the chiral connection, given by
\begin{equation}
\Gamma_{\mu}=\frac{1}{2}\left(u^{\dagger} \partial_{\mu} u 
+ u \partial_{\mu} u^{\dagger}\right) ,
\label{u}
\end{equation}
with 
\begin{align}
u &=\exp\left(\frac{ i}{\sqrt{2}F_{\pi}}\phi^{8}\right) , \label{eq:u}
\end{align}
where $F_{\pi}=92.1$ MeV is the pion decay constant~\cite{ParticleDataGroup:2022pth}, and $\phi^8$ 
is the pseudoscalar SU(3) octet, that is
\begin{align}
\phi^{8}=\left(\begin{array}{ccc}
\frac{\pi^{0}}{\sqrt{2}}+\frac{1}{\sqrt{6}} \eta_8 & \pi^{+} & K^{+} \\
\pi^{-} & -\frac{1}{\sqrt{2}} \pi^{0}+\frac{1}{\sqrt{6}} \eta_8 & K^{0} \\
K^{-} & \bar{K}^{0} & -\frac{2}{\sqrt{6}} \eta_8
\end{array}\right)  .
\label{pseudo}
\end{align}

In addition, the physical $\eta$ and $\eta'$ mesons are 
    the mixtures of $\eta^8$ and $\eta^{1}$
    \begin{align}
      \left(\begin{array}{c}
      \left|\eta\right\rangle \\
      \left|\eta^{\prime}\right\rangle
      \end{array}\right)=\left(\begin{array}{cc}
      -\sin \theta_{P} & \cos \theta_{P} \\
       \cos \theta_{P} & \sin\theta_{P}
      \end{array}\right)\left(\begin{array}{l}
      \left|\eta^1\right\rangle \\
      \left|\eta^8\right\rangle
      \end{array}\right) ,
      \label{eta}
      \end{align}
   where $\eta^1$ becomes the ninth pseudo-Goldstone boson in large $N_c$ 
    QCD \cite{tHooft:1973alw,Witten:1979vv,Veneziano:1980xs,Guo:2011pa}. The Goldstone boson nonet is written as 
    \begin{eqnarray}
      \phi^9 &=& \phi^8 + \frac{1}{\sqrt{3}} \eta^1,
    \end{eqnarray}
    which leads to a relation in the commutator
    \begin{eqnarray}
      \left[ \phi^9, \, \partial_{\mu}\phi^{9}\right] &=& \left[ \phi^8, \, \partial_{\mu}\phi^{8}\right].
    \end{eqnarray}
    Therefore, only the scattering of the octet Goldstone bosons  off 
    the axial-vector mesons in Weinberg-Tomozawa term 
    contributes to $J^{P\left(C\right)}=1^{-\left(+\right)}$ 
    spectrum.

The covariant derivative $\mathcal{D}_{\mu}$ by means 
of the connection $\Gamma_{\mu}$ encodes 
the leading order interaction between the pseudoscalar mesons 
and the vector field $V_{\mu}$~\cite{Ecker:1989yg,Meissner:1987ge,Birse:1996hd}. 
Therefore, by replacing the 
$V_{\mu}$ field to the axial-vector field $A_\mu$ corresponding to either the 
$A_1$ or $B_1$ multiplet, the chiral transition between $\phi^{8}$ 
(pseudoscalar) and $A\left(1^{+}\right)$ (axial-vector) is 
described by the following interaction Lagrangian
\begin{equation}
\mathcal{L}_{I}=-\frac{1}{4 F_{\pi}^{2}}\left\langle\left[A^{\mu}, \partial^{\nu} A_{\mu}\right]\left[\phi^{8}, \partial_{\nu} \phi^{8}\right]\right\rangle ,
\label{intlag}
\end{equation}
which accounts for the WT interaction 
term for the $PA \to PA$ process, with $P$ and $A$ 
corresponding to the pseudoscalar and axial-vector mesons, 
respectively. From this Lagrangian we obtain the 
$S$-wave transition amplitude among the channels listed 
in Tables~\ref{tab1}, \ref{tab2}, \ref{tab3} and \ref{tab4}, 
that is
\begin{align}
V_{i j}(s)= -\frac{\epsilon \cdot \epsilon^{\prime}}{8 F_{\pi}^{2}} C_{i j}\left[3 s-\left(M^{2}+m^{2}+M^{\prime^{2}}+m^{\prime 2}\right)-\frac{1}{s}\left(M^{2}-m^{2}\right)\left(M^{\prime 2}-m^{\prime 2}\right)\right] ,
\label{wt}
\end{align}
where $\epsilon\left(\epsilon^{\prime}\right)$ 
stands for the polarization four-vector of the 
incoming (outgoing) axial-vector meson~\cite{Borasoy:1995ds,Roca:2005nm}. 
The masses $M\left(M^{\prime}\right), m\left(m^{\prime}\right)$ 
correspond to the initial (final) axial-vector mesons 
and initial (final) pseudoscalar mesons, respectively. 
The indices $i$ and $j$ represent 
the initial and final $PA$ states, respectively. 
The coefficients $C_{ij}$ are given in Tables 
\ref{cij0}, \ref{cij1}, \ref{cijha}, and \ref{cijhb}.

\begin{table}[h!]
\caption{$C_{ij}$ coefficients in Eq.~\eqref{wt} for axial and pseudoscalar pairs 
coupled to $J^{PC}=1^{-+}$ in $S$-wave and $I=0$.}
\centering
\begin{tabular}{c | c c c c c}
\hline\hline
 $C_{ij}$~& ~$a_1\pi$~ & ~$K_1(1270)\bar{K}$~ & ~$f_1(1285)\eta$~ & ~$K_1(1400)\bar{K}$ ~&~ $f_1(1420)\eta$\\
\hline
$a_1\pi$ ~& $-4$ & $\sqrt{\frac{3}{2}}\sin \theta_{K_1}$ & $0$ & $\sqrt{\frac{3}{2}}\cos \theta_{K_1}$ & $0$\\
$K_1(1270)\bar{K}$ ~& & $-3$ & $-\frac{3}{\sqrt{2}}\sin \theta_{^3P_1}\sin {\theta_{K_1}}$ & $0$ & $-\frac{3}{\sqrt{2}}\,\cos \theta_{^3P_1}\,\sin {\theta_{K_1}}$\\
$f_1(1285)\eta$ ~& & & $0$ & $-\frac{3}{\sqrt{2}}\,\cos \theta_{K_1}\,\sin {\theta_{^3P_1}}$ & $0$\\
$K_1(1400)\bar{K}$ ~&  &  & & $-3$ & $-\frac{3}{\sqrt{2}}\cos \theta_{^3P_1}\cos \theta_{K_1}$\\
$f_1(1420)\eta$ ~& &  &  &  & $0$\\
\hline\hline
\end{tabular}
\label{cij0}
\end{table}

\begin{table}[h!]
\caption{$C_{ij}$ coefficients in Eq.~\eqref{wt} for axial and pseudoscalar pairs 
coupled to $J^{PC}=1^{-+}$ in $S$-wave and $I=1$.}
\centering
\begin{tabular}{c | c c c c c c}
\hline\hline
 $C_{ij}$~& ~$b_1\pi$~ & ~$f_1(1285)\pi$~ & ~$f_1(1420)\pi$ ~&~$K_1(1270)\bar{K}$ ~& ~$a_1\eta$ ~&~ $K1(1400)\bar{K}$\\
\hline
$b_1\pi$ ~& $-2$ & $0$ & $0$ & $\cos\theta_{K_1}$ & $0$ & $-\sin\theta_{K_1}$\\
$f_1(1285)\pi$ ~& & $0$ & $0$ & $\sqrt{\frac{3}{2}}\sin\theta_{K_1}\sin\theta_{^3P_1}$ & $0$ & $\sqrt{\frac{3}{2}}\cos\theta_{K_1}\sin \theta_{^3P_1}$\\
$f_1(1420)\pi$ ~& & & $0$ & $\sqrt{\frac{3}{2}}\cos\theta_{^3P_1}\sin\theta_{K_1}$ & $0$ & $\sqrt{\frac{3}{2}}\cos\theta_{K_1}\cos\theta_{^3P_1}$\\
$K_1(1270)\bar{K}$ ~& & & & $-1$ & $-\sqrt{\frac{3}{2}}\sin\theta_{K_1}$ & $0$\\
$a_1\eta$ ~& & & & & $0$ & $-\sqrt{\frac{3}{2}}\cos\theta_{K_1}$\\
$K_1(1400)\bar{K}$ ~& & & & & & $-1$\\
\hline\hline
\end{tabular}
\label{cij1}
\end{table}

\begin{table}[h!]
\caption{$C_{ij}$ coefficients in Eq.~\eqref{wt} for axial and pseudoscalar pairs 
coupled to $J^{P}=1^{-}$ in $S$-wave and $I=1/2$. Here the flavor-neutral axial mesons have $J^{PC}=1^{++}$.} 
\centering
\begin{tabular}{c | c c c c c}
\hline\hline
 $C_{ij}$~& ~$a_1 K$~ & ~$f_1(1285) K$~ & ~$K_1(1270) \eta$~ & ~$f_1(1420) K$ ~&~ $K_1(1400)\eta$\\
\hline
$a_1 K$ ~& $-2$ & $0$ & $-\frac{3}{2}\sin\theta_{K_1}$ & $0$ & $-\frac{3}{2}\cos\theta_{K_1}$\\
$f_1(1285) K$ ~& & $0$ & $\frac{3}{2}\sin\theta_{K_1}\sin\theta_{^3P_1}$ & $0$ & $\frac{3}{2}\sin\theta_{K_1}\cos\theta_{K_1}$\\
$K_1(1270) \eta$ ~& & & $0$ & $\frac{3}{2}\cos\theta_{^3P_1}\sin\theta_{K_1}$ & $0$\\
$f_1(1420) K$ ~& & & & $0$ & $\frac{3}{2}\cos\theta_{^3P_1}\cos\theta_{K_1}$\\
$K_1(1400)\eta$ ~& & & & & $0$\\
\hline\hline
\end{tabular}
\label{cijha}
\end{table}

\begin{table}[h!]
\caption{$C_{ij}$ coefficients in Eq.~\eqref{wt} for axial and pseudoscalar pairs 
coupled to $J^{P}=1^{-}$ in $S$-wave and $I=1/2$. Here the flavor-neutral axial mesons have $J^{PC}=1^{+-}$.}
\centering
\begin{tabular}{c | c c c c c}
\hline\hline
 $C_{ij}$~& ~$h_1(1170) K$~ & ~$b_1 K$~ & ~$K_1(1270) \eta$~ & ~$h_1(1415) K$ ~&~ $K_1(1400)\eta$\\
\hline
$h_1(1170) K$ ~& $0$ & $0$ & $\frac{3}{2}\cos\theta_{K_1}\sin\theta_{^1P_1}$ & $0$ & $\frac{3}{2}\sin\theta_{K_1}\sin\theta_{^1P_1}$\\
$b_1 K$ ~& & $-2$ & $-\frac{3}{2}\cos\theta_{K_1}$ & $0$ & $-\frac{3}{2}\sin\theta_{K_1}$\\
$K_1(1270) \eta$ ~& & & $0$ & $\frac{3}{2}\cos\theta_{K_1}\cos\theta_{^1P_1}$ & $0$\\
$h_1(1415) K$ ~& & & & $0$ & $\frac{3}{2}\sin\theta_{K_1}\cos\theta_{^1P_1}$\\
$K_1(1400)\eta$ ~& & & & & $0$\\
\hline\hline
\end{tabular}
\label{cijhb}
\end{table}

Before proceeding, let us discuss the $A_1\phi^8\to B_1\phi^8$ transitions, with $A_1 $ and $B_1$ the two SU(3) octets of axial-vector mesons and $\phi^8$ the octet of the pseudo-Nambu-Goldstone bosons.
Let $A_{1\mu}$ and $B_{1\mu}$ denote the fields for the $1^{++}$ and $1^{+-}$ axial-vector meson multiplets, respectively. Under parity transformation, we have $A_{1\mu} \to -A_{1}^{\mu}$ and $B_{1\mu} \to -B_{1}^{\mu}$; under charge conjugation, we have $A_{1\mu} \to A_{1\mu}^T$ and $B_{1\mu} \to -B_{1\mu}^T$. Then the $A_1\phi^8\to B_1\phi^8$ transitions can only arise at $\mathcal{O}\left(p^2\right)$ with $p$ the momentum scale in the chiral power counting. They are given by operators such as 
\begin{equation}
    \left\langle A_{1\mu} [B_{1\nu}, [u^\mu,u^\nu]] \right\rangle, \label{eq:a1b1}
\end{equation}
with 
\begin{equation}
    u_\mu=i\left(u^{\dagger}\partial_\mu u-u\partial_\mu u^{\dagger}\right).\label{eq:umu}
\end{equation}
Such terms are one order higher in the chiral power counting than the WT terms describing the $A_1\phi^8\to A_1\phi^8$ and $B_1\phi^8\to B_1\phi^8$ transitions, and thus will be neglected.

\subsection{Unitarization procedure}
\label{sec:unit}

The unitarization procedure we adopt follows ChUA in which the scattering 
amplitudes in Eq.~\eqref{wt} are the elements of a matrix, 
denoted by $V$, such that it enters as an input to solve 
the Bethe-Salpeter equation, which in its on-shell 
factorization form, reads~\cite{Oller:2000ma}
\begin{equation}
T=(1-V\,G)^{-1}\, V \, .
\label{bs}
\end{equation}
The $V$ matrix describes the transition 
between the channels listed in Tables~\ref{tab1}, \
\ref{tab2}, \ref{tab3}, and \ref{tab4}. In addition, 
$G$ is the diagonal loop function 
matrix whose diagonal matrix elements are given by
\begin{equation}
G_{l}=i \int \frac{d^{4} q}{(2 \pi)^{4}}
 \frac{1}{q^{2}-m^{2}_l+i \epsilon} \frac{1}{(q-P)^{2}-M^{2}_l+i \epsilon} \, ,
\label{loop}
\end{equation}
with $m_l$ and $M_l$ the masses of the pseudoscalar 
and axial-vector mesons, respectively, involved in the 
loop in the channel $l$, and $P$ the total 
four-momentum of those mesons ($P^2=s$). After the integration 
over the temporal component $q^{0}$, Eq.~\eqref{loop} becomes
\begin{equation}
G_{l}(s)=\int \frac{d^{3} q}{(2 \pi)^{3}} \frac{\omega_{1}+\omega_{2}}{2 \omega_{1} \omega_{2}} \frac{1}{\left(P^{0}\right)^{2}-\left(\omega_{1}+\omega_{2}\right)^{2}+i \epsilon}\, ,
\label{loopcut}
\end{equation}
with $\omega_{1}=\sqrt{{M_l}^{2}+|\vec{q}|^{2}}$ and 
$\omega_2 = \sqrt{{m_l}^{2}+|\vec{q}|^{2}}$, and can be regularized 
by means of a cutoff in the three-momentum 
$q_{\max }$. On the other hand, the function $G_{l}$ can also 
be regularized using a subtraction constant as~\cite{Oller:1998zr} 
\begin{align}
G^{\rm DR}_{l}(s)=& \frac{1}{16 \pi^{2}}\left[\alpha_{l}(\mu)+\log \frac{M_{l}^{2}}{\mu^{2}}+\frac{m_{l}^{2}-M_{l}^{2}+s}{2 s} \log \frac{m_{l}^{2}}{M_{l}^{2}}\right.
+\frac{p_l}{\sqrt{s}}\left(\log \frac{s-m_{l}^{2}+M_{l}^{2}+2 p_l \sqrt{s}}{-s+m_{l}^{2}-M_{l}^{2}+2 p_l \sqrt{s}}\right.\nonumber\\
&\left.\left.+\log \frac{s+m_{l}^{2}-M_{l}^{2}+2 p_l \sqrt{s}}{-s-m_{l}^{2}+M_{l}^{2}+2 p_l \sqrt{s}}\right)\right] \, ,
\label{loopdr}
\end{align}
where $p_l$ is the three-momentum of the mesons in the 
center-of-mass (c.m.) frame
\begin{equation}
p_l=\frac{\sqrt{\left(s-\left(M_{l}+m_{l}\right)^{2}\right)\left(s-\left(M_{l}-m_{l}\right)^{2}\right)}}{2 \sqrt{s}} \, ,
\end{equation}
while $\mu$ is an arbitrary scale of the regularization. 
Any changes in the $\mu$ scale can be absorbed by 
the subtraction constant $\alpha_l(\mu)$ such that the 
result is independent of the scale. We may
determine the subtraction constant for each intermediate 
state of the scattering problem by comparing Eqs.~\eqref{loopcut}, regularized using $q_\text{max}$, 
and \eqref{loopdr} at the threshold. The equivalence 
between the two prescriptions for the loop-function is 
discussed in, e.g., Refs.~\cite{Oller:1998hw,Oller:2000fj,Fu:2021wde}. In this work, we follow 
Ref.~\cite{Geng:2008gx} and 
set $\mu=1$~GeV and $\alpha=-1.35$, which 
is obtained by matching to hard cutoff regularization with $q_\text{max}\simeq 0.7$~GeV in the $f_1(1285)\eta$ channel. 
This set of parameters are used for all channels, and a variation of the cutoff within $q_\text{max}= (0.7\pm0.1)$~GeV, and correspondingly $\alpha(\mu=1\,\text{GeV}) = -1.35\pm {0.17}$, will be used to show the dependence of the results on this parameter. 

\subsection{Searching for poles}
\label{subsec:poles}

We move on to the complex energy plane to search 
for poles in the 
$T$-matrix. Specifically, for a single-channel problem, there are two Riemann sheets for the complex energy plane. Bound states show up as poles, 
below the threshold, in the transition matrix on the real energy axis on the first Riemann 
sheet, while virtual states manifest 
themselves below the threshold on the real axis on the second Riemann sheet, and resonances correspond to poles off the real axis on the second Riemann sheet. 
The Riemann sheets come about because the $G$ loop function 
has a cut extending from the threshold to infinity which is usually chosen to be along the positive real axis. For $n$ coupled channels, there are $n$ cuts and thus $2^n$ Riemann sheets. From unitarity and the Schwarz reflection principle, the 
discontinuity of the $G_l$ function can be read off from its imaginary part,
\begin{equation}
\textrm{Im}\, G_l(s) = -\frac{p_l}{8\pi\sqrt{s}}\, , 
\end{equation}
which we can use to perform an analytic continuation 
to the entire complex plane. In this case, the $G_l$ 
loop function on the ``second'' Riemann sheet with respect to the $l$th channel reads
\begin{equation}
G^{\rm II}_l(s) = G^{\rm I}_l(s) 
+ i\,\frac{p_l}{4\pi\sqrt{s}}\,; 
\end{equation}
the lower half plane of this Riemann sheet is directly connected to the physical region when the $l$th channel is open, {\it i.e.}, $\textrm{Re}(\sqrt{s})\geq m+M$.
We will label the Riemann sheets according to the sign of the imaginary part of the corresponding c.m. momentum for each channel (see the next section).

Furthermore, it is also possible to determine 
the pole couplings to the $l$th channel. Note 
that close to the pole singularity the $T$-matrix 
elements $T_{ij}(s)$ admit a Laurent expansion,  
\begin{equation}
T_{ij}(s)= \frac{g_i\,g_j}{s-z_p} + \text{regular terms} ,
\end{equation}
where $z_p=(M_p - i\Gamma/2)^2$ is the pole location on 
the complex energy plane, with $M_p$ and $\Gamma$ 
standing for the pole mass and width, respectively. 
Therefore, the product of couplings $g_i g_j$ is the residue at the 
pole in 
$T_{ij}(s)$ which takes values on the Riemann sheet where the pole is located. In this way, the couplings can be evaluated straightforwardly. For instance, 
for a diagonal transition it is given by
\begin{align}
g_{i}^{2}&=\frac{r}{2 \pi} \int_{0}^{2 \pi} 
T_{i i}(z(\theta)) e^{i \theta} d \theta \notag\\
&= \lim_{s\to z_p} (s-z_p) T_{ii}(s) = \left[ \frac{d}{ds} \frac1{T_{ii}(s)} \right]^{-1}_{s=z_p} \, ,
\end{align}
where $z(\theta) = z_p + \,re^{i\theta}$ with $r$ the radius of contour for the integral, and the two lines give two equivalent ways of computing residues.

\section{$\eta_1(1855)$ and its decays}
\label{sec:eta}

\subsection{Dynamical generation of the $\eta_1(1855)$}
\label{subsec:gen}

Following the unitarization procedure described 
previously, we seek dynamically generated states 
stemming from the $S$-wave interactions between pseudoscalar and axial-vector mesons. 
For the $I=0$ case, the transition amplitudes among the 
channels listed in Table~\ref{tab1} are determined using 
Eq.~\eqref{wt} with the $C_{ij}$ coefficients given in 
Table~\ref{cij0}. In Table \ref{etapoles}, we show the isoscalar poles with exotic quantum numbers $J^{PC}=1^{- +}$
obtained by solving Eq.~\eqref{bs} using those coefficients 
as well as each set of mixing angles listed in Table \ref{angles}.
We also show the couplings of these poles to the channels 
spanning the space of states in Table~\ref{tab1}.

{\squeezetable \begin{table}[tb]
\caption{The poles (in GeV) and their corresponding couplings (in GeV) to 
the channels contributing to the $PA$ interaction with $I=0$ and exotic quantum numbers $J^{PC}=1^{-+}$. The corresponding Riemann sheet for each pole is listed below the pole position. The dominantly coupled channel is emphasized in boldface for each pole. The errors of the poles are from varying the subtraction constant within $\alpha(\mu=1\,\text{GeV})=-1.35\pm0.17$, and only the central values of the couplings are given.
For each pole, we also give the central values of the peak mass and width as $(M_\text{peak},\Gamma_\text{peak})$, after considering the axial-vector meson widths, in the last row of the corresponding panel.
} 
 \centering
  \begin{tabular}{c c c c c c }
  \hline\hline
  {Poles} (Set A) & &~~~~~~~~ {Channels} & & & \\
  \hline\hline
  $\bm{1.39\pm0.01-i (0.04\pm0.01)}$ & $a_1\pi$ ~&~ $K_1(1270) \bar{K}$ ~&~ $f_1(1285)\eta$ ~&~ $K_1(1400) \bar{K}$ ~&~ $f_1(1420)\eta$\\
  $(-++++)$ & & & & & \\
  $g_l$ & $\bm{5.21+i3.01 }$ & $1.22+i0.78$ & $0.01+i0.02$ & $0.36+i0.35$ & $0.00$\\
    $\bm{(1.39, 0.24)}$ &  &  &  &  &  \\
  \hline\hline
  $\bm{1.69\pm0.03}$ ~&~ $a_1\pi$ ~&~ $K_1(1270) \bar{K}$ ~&~ $f_1(1285)\eta$ ~&~ $K_1(1400) \bar{K}$ ~&~ $f_1(1420)\eta$\\
  $(-++++)$ & & & & & \\
  $g_l$ & $0.36+i0.98$ & $\bm{8.16-i0.17}$ & $3.64+i0.01$ & $0.09-i0.15$ & $2.46+i0.01$\\
    $\bm{(1.69, 0.08)}$ &  &  &  &  &  \\
  \hline\hline
  $\bm{1.84\pm0.03}$ ~&~ $a_1\pi$ ~&~ $K_1(1270) \bar{K}$ ~&~ $f_1(1285)\eta$ ~&~ $K_1(1400) \bar{K}$ ~&~ $f_1(1420)\eta$\\
  $(---++)$ & & & & & \\ 
  $g_l$ & $0.07+i0.28$ & $0.69+i0.55$ & $1.68+i0.08$ & $\bm{9.33+i0.15}$ & $1.16+i0.06$\\
    $\bm{(1.84, 0.16)}$ &  &  &  &  &  \\
  \hline\hline
  {Poles} (Set B) & &~~~~~~~~ {Channels} & & & \\
  \hline
  $\bm{1.39\pm0.01-i (0.04\pm0.01)}$ ~&~ $a_1\pi$ ~&~ $K_1(1270) \bar{K}$ ~&~ $f_1(1285)\eta$ ~&~ $K_1(1400) \bar{K}$ ~&~ $f_1(1420)\eta$\\
  $(-++++)$ & & & & & \\
  $g_l$ & $\bm{5.21+i3.03}$ & $0.81+i0.53$ & $0.00$ & $0.55+i0.54$ & $0.00$\\
    $\bm{(1.42, 0.34)}$ &  &  &  &  &  \\
    \hline\hline
  $\bm{1.70\pm0.02}$  ~&~ $a_1\pi$ ~&~ $K_1(1270) \bar{K}$ ~&~ $f_1(1285)\eta$ ~&~ $K_1(1400) \bar{K}$ ~&~ $f_1(1420)\eta$\\
  $(-++++)$ & & & & & \\
  $g_l$ & $0.25+i0.67$ & $\bm{8.34-i0.08}$ & $1.27-i0.01$ & $0.37+i0.17$ & $2.58-i0.01$\\
    $\bm{(1.70, 0.10)}$ &  &  &  &  &  \\
  \hline\hline
  $\bm{1.84\pm0.03}$  ~&~ $a_1\pi$ ~&~ $K_1(1270) \bar{K}$ ~&~ $f_1(1285)\eta$ ~&~ $K_1(1400) \bar{K}$ ~&~ $f_1(1420)\eta$\\
  $(---++)$ & & & & & \\
  $g_l$ & $0.15+i0.62$ & $0.33-i0.27$ & $1.83+i0.09$ & $\bm{9.05+i0.17}$ & $3.81-i0.20$\\
    $\bm{(1.85, 0.18)}$ &  &  &  &  &  \\
  \hline\hline  
  \end{tabular}
  \label{etapoles}
  \end{table}}

Furthermore, in Table \ref{etapoles} we also highlight 
the Riemann sheets, the first and the second one for each channel, denoted 
by the $+$ and $-$ signs, respectively. We get three poles such 
that their locations are barely affected by the change of the mixing angles from set A to set B listed in Table~\ref{angles}. The lower pole is at $1.39$~GeV with a width 
of about $0.04$~GeV, which is above the $a_1\pi$ threshold. In particular, 
this channel is open for decay, and the fact that it is this channel the 
one for which the pole couples mostly, as pointed out in Table~\ref{etapoles}, 
explains why that pole has such a value for its width. By contrast, 
although the $a_1\pi$ channel is also open for decay, the pole 
at $1.69$~GeV has a much smaller width because its coupling to this channel 
is small compared to the one for $K_1(1400)\bar{K}$, which 
is the dominant channel for that pole. Similarly, the highest pole, 
located at $1.84$~GeV, couples mostly to the $K_1(1400) \bar{K}$ channel, 
and has a small imaginary part.
In addition, we can also understand why the highest pole couples more 
to the $K_1(1400)\bar{K}$ than to the $f_1(1285)\eta$. The latter channel 
is closer to the pole than the former, but from Table \ref{cij0}, the diagonal 
$f_1(1285)\eta$ transition is not allowed since its WT term is 
zero. Nevertheless, the pole couples to $f_1(1285)\eta$ through the nondiagonal 
$K_1(1400) \bar{K}$--$f_1(1285)\eta$ transition, which leads to a small coupling.

\subsection{Effects of the widths of the axial-vector mesons}
\label{sec:widtheffects}

So far we have neglected the nonzero widths of the axial-vector mesons. In order to investigate their effects on the results, we use complex masses for the intermediate resonances, that is, $M_i\to M_i-i\Gamma_i/2$. However, by doing that, the analytic properties are lost such that the poles of the $T$ matrix do not correspond to the masses and widths of the obtained resonances any more. On the other hand, we can see the impact of such nonzero widths on the lineshapes of the transition matrix elements. 

\begin{figure}[tb]
  \includegraphics[width=80mm]{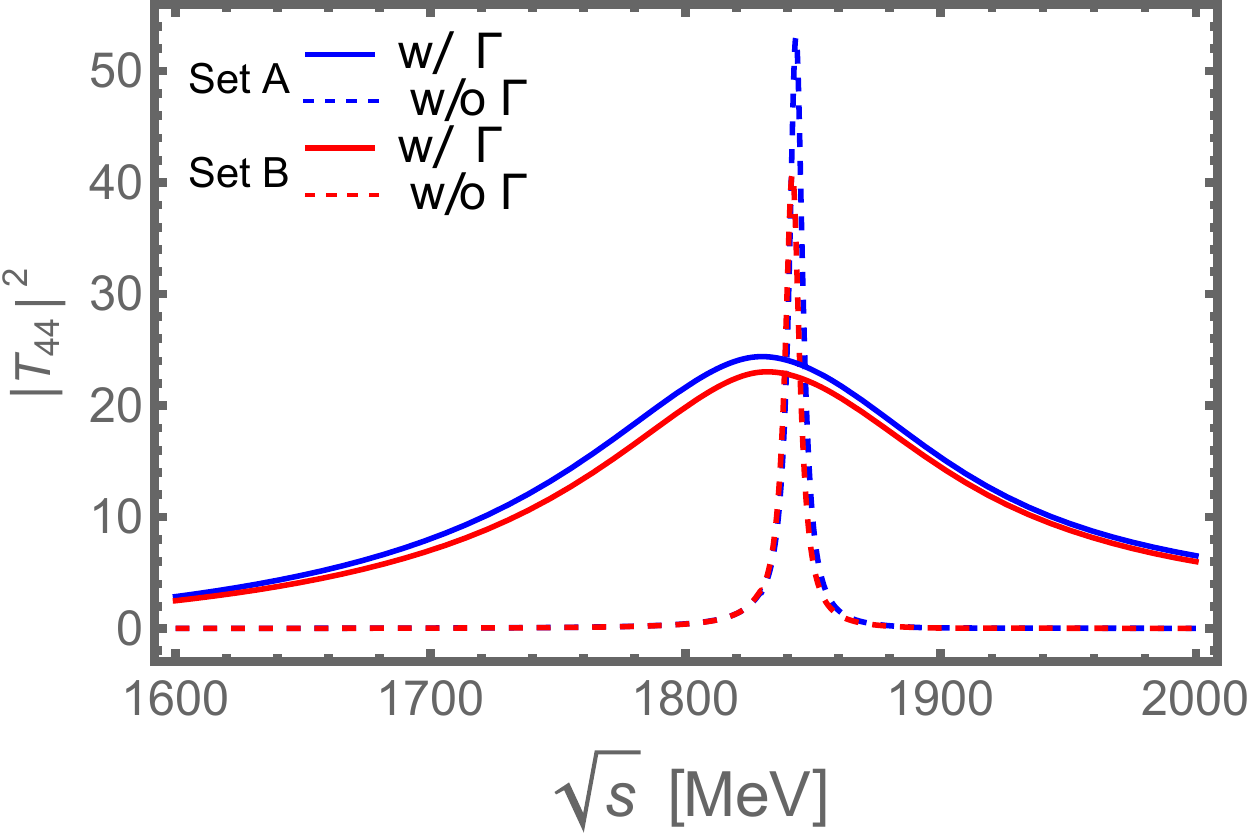}  
  \caption{The blue dashed and solid lines are, respectively, the modulus squared of the $T$-matrix element, corresponding to the diagonal $K_1(1400)\bar{K}\to K_1(1400)\bar{K}$ transition, evaluated with and without the inclusion of the widths associated with the axial-vector mesons taking part in the loop function $G_l$ (Eq.~\eqref{loop}).}
  \label{fig:isoscalar33BW}
\end{figure}

In Fig.~\ref{fig:isoscalar33BW} we show a comparison between the lineshape for the $T$-matrix element corresponding to the elastic transition $T_{K_1(1400)\bar{K}\to K_1(1400)\bar{K}}$ with and without including the widths for the intermediate particles. This channel has the strongest coupling to the pole at $1.84$ GeV; therefore, we expect that any nontrivial structure should manifest most in its associated $T$-matrix element. The dashed and  solid lines are the $T_{K_1(1400)\bar{K}\to K_1(1400)\bar{K}}$ with zero and nonzero width, respectively, for both sets $A$ and $B$ of mixing angles in Table~\ref{fig:isoscalar33BW}. Notice that, for the case of zero width approximation, the $T_{K_1(1400)\bar{K}\to K_1(1400)\bar{K}}$ lineshape has narrow peaks around $1845$ MeV, right at the range of energy where we expect the $\eta_1(1855)$ manifests in our model. 
The inclusion of finite widths for the axial-vector mesons changes the sharp peak to a broad bump with a width of about 0.2~GeV, which is around the width of the $K_1(1400)$~\cite{ParticleDataGroup:2022pth}. 
Notice that the width matches nicely that of the $\eta_1(1855)$ measured by BESIII, $\left(188 \pm 18_{-8}^{+3}\right) \mathrm{MeV}$~\cite{BESIII:2022riz}.
One can obtain a peak mass $M_\text{peak}$ and a peak width $\Gamma_\text{peak}$, defined as the width at the half maximum of the line shape of the diagonal $T$-matrix element modulus squared for the dominant channel. In Table~\ref{etapoles}, we also list the values of $M_\text{peak}$ and $\Gamma_\text{peak}$ for each pole. 
The values for the highest $\eta_1$ pole in the table agree remarkably well with those of the $\eta_1(1855)$.
These values may be compared with the resonance parameters extracted using the Breit-Wigner form in experimental analyses.

\begin{figure}[tb]
    \centering
    \includegraphics[width=0.45\textwidth]{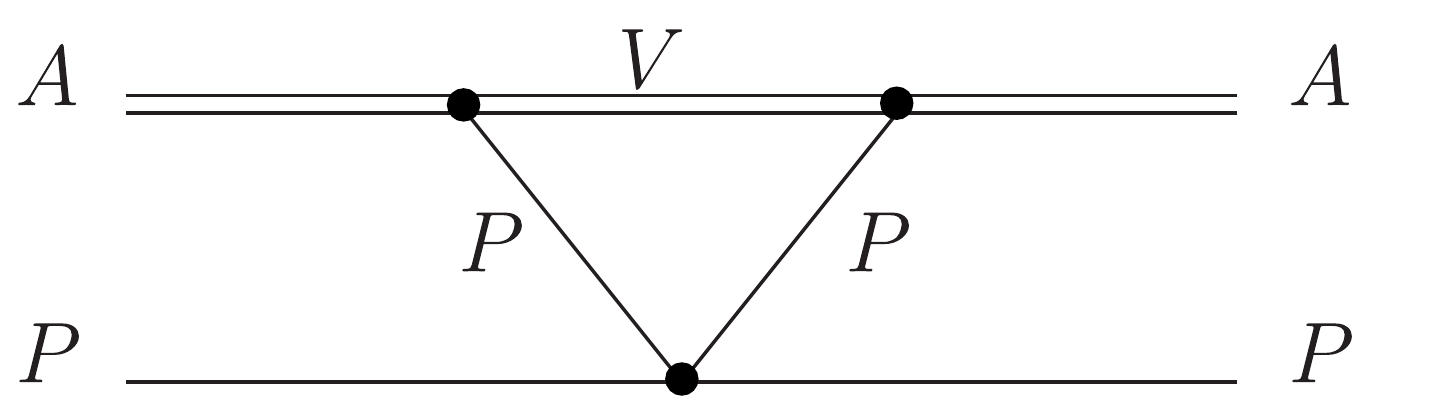} \hfill
    \includegraphics[width=0.45\textwidth]{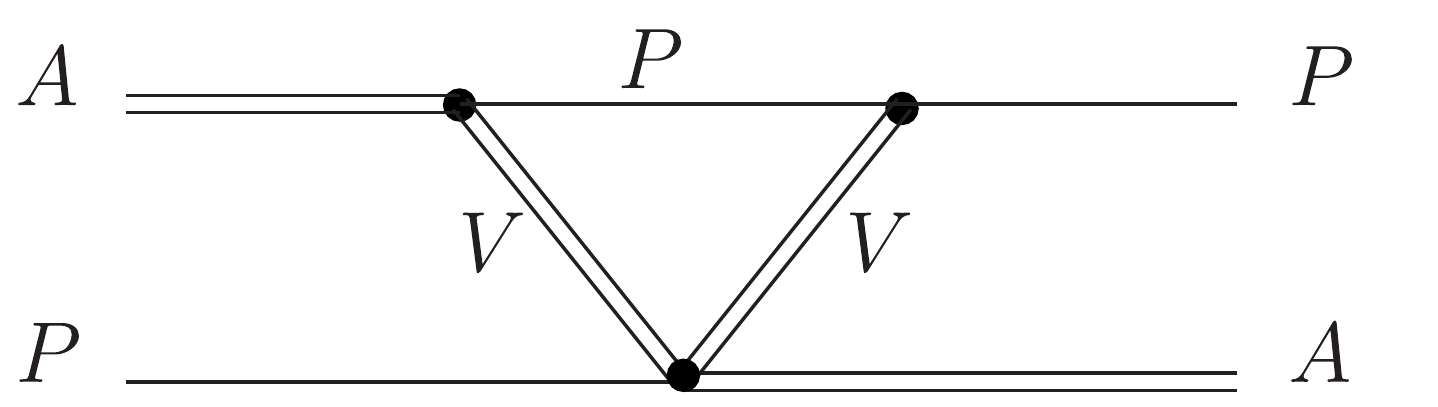}
    \caption{Possible triangle diagram contributions to the scattering between an axial vector meson and a pseudoscalar meson.}
    \label{fig:AVPimprove}
\end{figure}

The results presented may be  improved by considering one additional possible contribution due to the axial-meson decays, shown in Fig.~\ref{fig:AVPimprove}.
Since the intermediate vector and pseudoscalar mesons in the triangle diagrams can go on shell, such contributions could further increase the widths of the dynamically generated states.
In the following, we will continue to present predictions neglecting the width effects of the axial-vector mesons.

Let us briefly discuss the other two predicted isoscalar exotic $\eta_1$ mesons in Table~\ref{etapoles}. The one with a mass of about 1.39~GeV, denoted as $\eta_1(1400)$, is expected to be rather broad due to the large width of the $a_1(1260)$ as it couples most strongly to the $a_1\pi$ channel. It can be searched for in final states such as $\rho\pi\pi$ and $K\bar K\pi\pi$.
The one with a mass around 1.7~GeV, denoted as $\eta_1(1700)$, couples most strongly to the $K_1(1270)\bar K$ and is expected to have a width similar to that of the $K_1(1270)$, {\it i.e.}, around 0.1~GeV.
It can also be searched for in final states of $K\bar K\pi\pi$.

\subsection{The $\eta_1(1855)\to \eta^{\prime}\eta$ 
and $K^*\bar{K}\pi$ decays}
\label{subsec:etadecay}

\begin{figure}[tbh]
\includegraphics[width=0.50\textwidth]{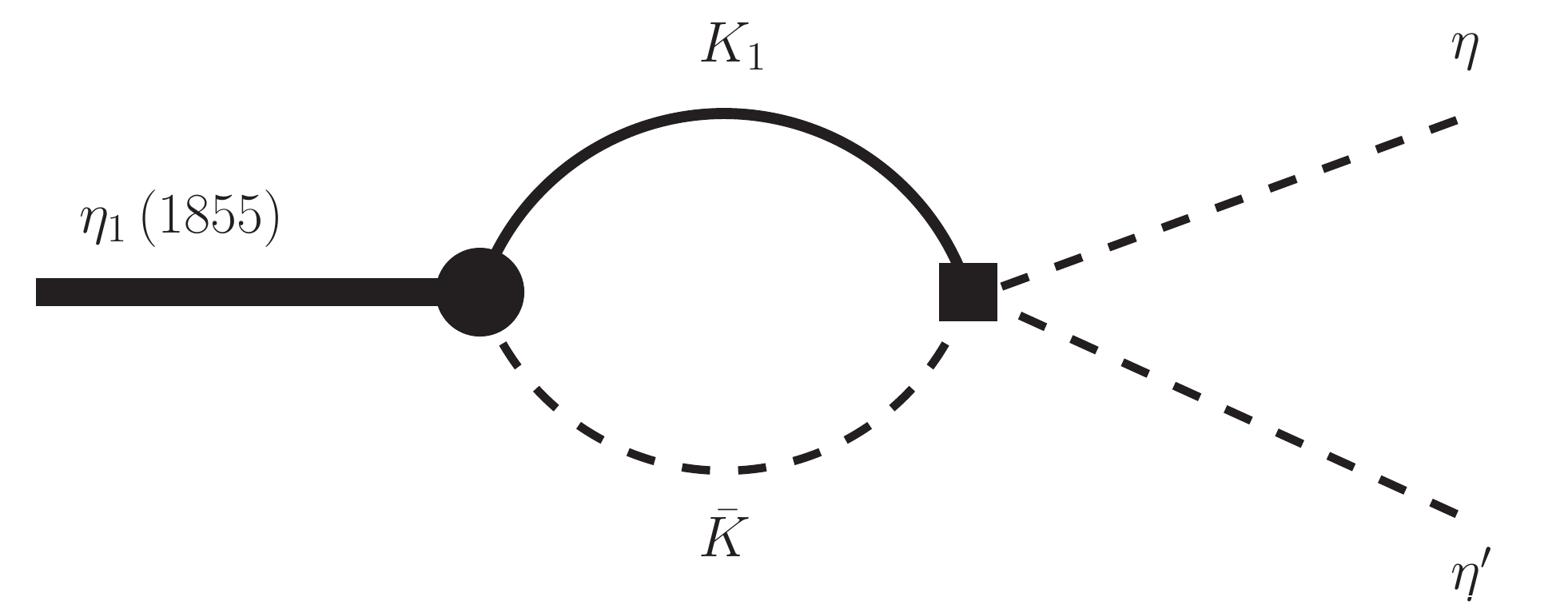}
\caption{Diagram corresponding to the $\eta_1\to \eta\eta^{\prime}$ decay through the $K_1\bar{K}$ loop.
}\label{gaeta}
\end{figure}

Let us first discuss the $\eta_1\to \eta\eta^{\prime}$ decay, 
whose Feynman diagram is shown in Fig.~\ref{gaeta}. Within our 
approach the $\eta_1(1855)$ structure decays via its $K_1(1400)\bar{K}$ 
component,  with the corresponding coupling constant listed in 
Table~\ref{etapoles}. We also need to evaluate 
the $K_1(1400)\bar{K}\to \eta\eta^{\prime}$ transition, for which we use the resonance chiral theory (R$\chi$T) operators given in Ref.~\cite{Cirigliano:2006hb}. 

The R$\chi$T operators can be divided regarding the intrinsic-parity sector to which they contribute. Due to its nature, the odd-intrinsic parity sector will contain a Levi-Civita tensor \cite{Wess:1971yu,Witten:1983tw,Kampf:2011ty}; for the $\eta_1\to\eta\eta'$ decay one cannot saturate the Lorentz indices in such tensor to get a nonzero contribution. Thus, only the even-intrinsic parity operators must give a nonvanishing contribution. Since the chiral $\mathcal{O}(p^2)$ Lagrangian does not contribute to such processes \cite{Ecker:1988te}, we will use the $\mathcal{O}(p^4)$ Lagrangian given in Ref. \cite{Cirigliano:2006hb}. From these operators, only three will contribute to this decay. To get the largest possible contribution from such operators, we use the upper bounds imposed from chiral counting as done in Ref. \cite{Miranda:2020wdg}. This amounts to making equal the three coupling constants and setting them to $\lambda_1^A=\lambda_2^A=\lambda_3^A=g=0.025\text{ GeV}^{-1}$, which gives a Lagrangian

\begin{eqnarray}
\mathcal{L}&=& g \left[\langle A_{\mu\nu}\left(u^{\mu}u_{\alpha}h^{\nu\alpha} + h^{\nu\alpha}u_{\alpha}u^{\mu}\right)\rangle + \langle A_{\mu\nu} \left( u_{\alpha}u^{\mu} h^{\nu\alpha}+ h^{\nu\alpha}u^{\mu}u_{\alpha}\right) \rangle + \frac{}{}\hspace*{-.5ex}\langle A_{\mu\nu}\left( u^{\mu}h^{\nu\alpha}u_{\alpha} + u_{\alpha} h^{\nu\alpha}u^{\mu}\right) \rangle \right],\nonumber\\\label{RChiTLag}
\end{eqnarray}
where $u_\mu$ has been given in Eq.~\eqref{eq:umu},
$h^{\mu\nu}=\mathcal{D}^{\{\mu} u^{\nu\}}$ is the symmetrized covariant derivative of $u_\mu$ and the spin-1 resonance field is given in the antisymmetric tensor formalism \cite{Ecker:1989yg}. However, since the $\eta_1\to K_1 \bar{K}$ transition is given in terms of Proca fields, we need to express the $K_1$ as a  Proca field. Following Ref. \cite{Ecker:1988te},
the antisymmetric tensor field can be expressed in terms of the Proca one as follows,
\begin{equation}
 R_{\mu}=\frac{1}{M_R}\partial^\nu R_{\nu\mu},
\end{equation}
where $M_R$ is the mass of the resonance. Using the Lagrangian of Eq.(\ref{RChiTLag}) and expressing the axial resonance in the Proca representation, we get the $\eta_1\to\eta\eta'$ decay amplitude 
\begin{equation}
\mathcal{M}_{\eta_1\to\, \eta\eta^{\prime}}=-\frac{4m_{\eta_1}^2}{3F_{\pi}^3 m_{K_1}}g g_{K_1(1400)\bar{K}} G_{K_1\bar{K}} \left[\left(  \alpha p_{\eta'}^2+\frac{1}{\sqrt{2}} \beta p_{\eta}^2\right)\varepsilon_{\eta_1}\cdot p_{\eta} + \left( p_{\eta}\leftrightarrow p_{\eta^{\prime}}\right) \right], \label{gaeta1}
\end{equation}
where $F_\pi$ is the pion decay constant, 
$g_{K_1\bar{K}}$ is the coupling constant of the pole to the $K_1(1400)\bar{K}$ channel, $G_{K_1\bar{K}}$ is the loop function for the $K_1$ and $\bar{K}$ mesons , $\varepsilon_{\eta_1}$ is the $\eta_1$ vector polarization, and $p_{\eta^{(\prime)}}$ is the momentum of the $\eta^{(\prime)}$. Here, $\alpha$ and $\beta$ are given in terms of the $\eta$-$\eta'$ mixing angle
\begin{subequations}
\begin{align}
 \alpha&=\cos{2\theta_P}+2\sqrt{2}\sin{2\theta_P},\\
 \beta&=2\sqrt{2}\cos{2\theta_P}-\sin{2\theta_P}.
\end{align}
\end{subequations}

Although one might try to rely in a much simpler way to describe the direct coupling of one axial-vector and three pseudoscalar fields by means of the hidden local symmetry (HLS) Lagrangian~\cite{Bando:1987ym,Bando:1987br,Kaiser:1990yf}, it is worth to notice that nonetheless, the total amplitude for this process given by the HLS Lagrangian vanishes, which coincides with Eq.(\ref{gaeta1}) in the chiral limit.

The decay of $\eta_1$ state into $\eta \eta^{\prime}$ is given 
by
\begin{equation}
\Gamma_{2B} = \frac{1}{2J+1}\,\frac{1}{8\pi M^2_{\eta_1}}\,\vert 
\mathcal{M}_{\eta_1\to\, \eta\eta^{\prime}}\vert^2\, q \, ,
\end{equation}
with the amplitude $\mathcal{M}_{\eta_1\to\, \eta\eta^{\prime}}$ 
in Eq.~\eqref{gaeta1}, while $J$ stands for the $\eta_1$ spin. Besides that, 
$q$ reads 
\begin{equation}
q=\frac{1}{2M_{\eta_1}}\lambda^{1/2}\left(M^2_{\eta_1},m^2_{\eta^{\prime}},m^2_{\eta}\right) , 
\end{equation}
with $M_{\eta_1}$, $m_{\eta^{\prime}}$, and $m_{\eta}$ the masses 
for the $\eta_1(1855)$, $\eta^{\prime}$, and $\eta$ mesons, respectively, where $\lambda\left(x,y,z\right)=x^2+y^2+z^2 -2xy -2yz-2zx$ is the K\"all\'en triangle function. 
Therefore, we get the following results for the decay width 
in this channel
  \begin{equation}
  \Gamma_{2 B}=\left\{\begin{array}{l}
    (19\pm4)\,\textrm{MeV (set A)}\,, \\
    (7\pm2) \,\textrm{MeV (set B)}\, ,
  \end{array}\right.
  \end{equation}
  where the error is from choosing subtraction constant to be in the range $\alpha(\mu=1 \mathrm{GeV})=-1.35 \pm 0.17$, corresponding to the hard cutoff $q_\text{max}=(0.7\pm 0.1)\,\rm{GeV}$ as discussed at the end of Section~\ref{sec:unit}.
For set $A$, our result agrees with that of Ref.~\cite{Dong:2022cuw}, where the $\eta_1(1855)$ was assumed to be a $K_1\bar K$ molecule and the same $\theta_{K_1}$ mixing angle was used for accounting for the 
$K_{1A}$ and $K_{1B}$ mixture contributing to the physical $K_1(1270)$ and 
$K_1(1400)$ states.

\begin{figure}[tbh]
\includegraphics[width=0.4\textwidth]{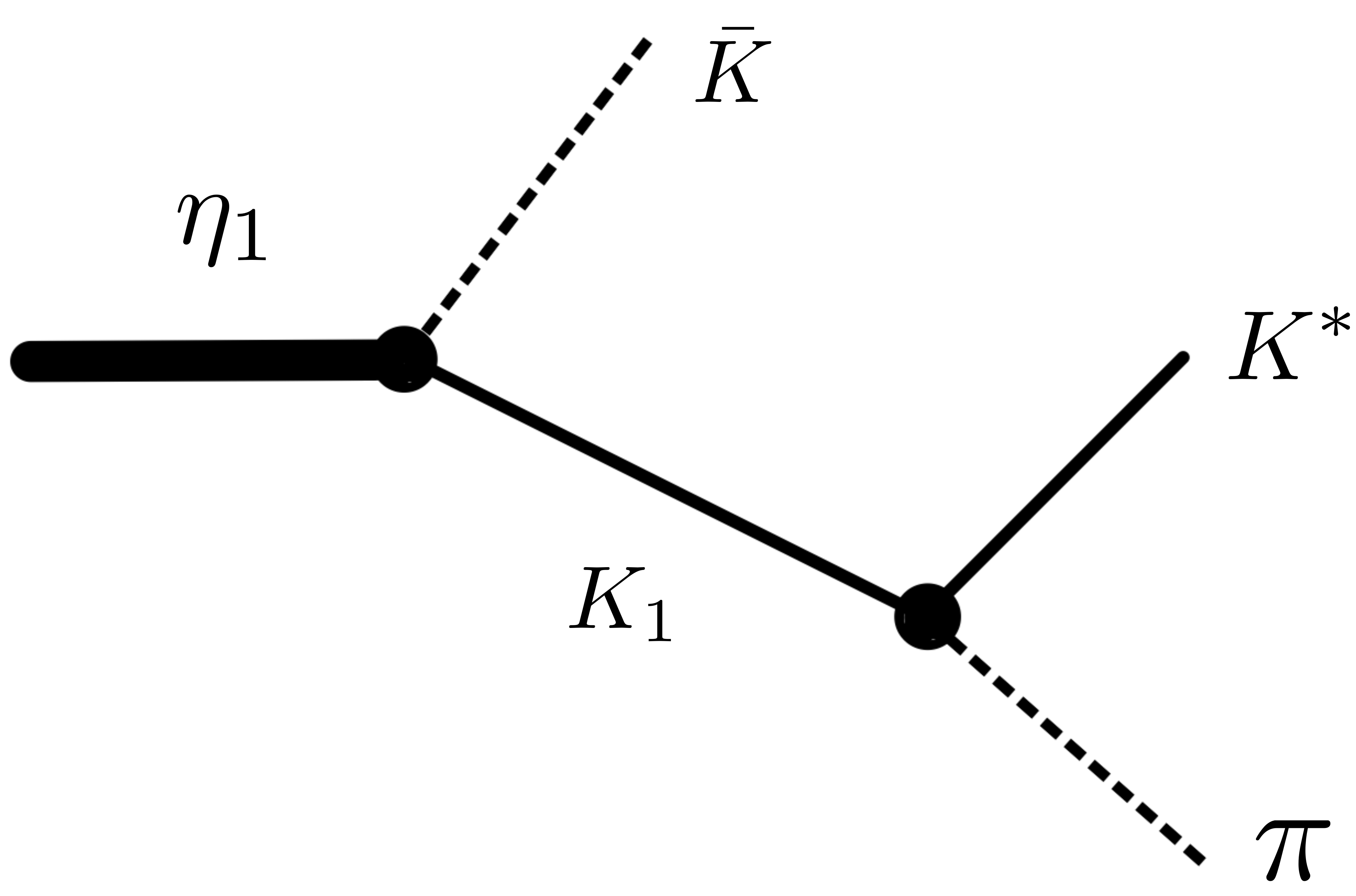}
\caption{Feynman Diagram associated with the three-body decay of 
the pole through its main component $K_1 \bar K$.}\label{gakk}
\end{figure}

As for the $\eta_1 \to \bar{K} K^*\pi$ three-body decay, 
Fig.~\ref{gakk} shows the Feynman diagrams contributing 
to this process. In particular, the $\eta_1(1855)$ decays through its 
molecular components, that in our approach are the 
$K_1(1270)\bar{K}$ and $K_1(1400)\bar{K}$. In this case, 
the contribution from the $K_1(1270)\bar{K}$ 
component can be ignored for the following reasons: 1) from Table~\ref{etapoles}, we see that 
the relative coupling strength for the $K_1(1270)\bar{K}$ 
channel is much smaller than that for the $K_1(1400)\bar{K}$ 
one; 2) the branching ratio $\mathcal{B}[K_1(1270) \to K^* 
\pi]$ is only $16\%$, while $96\%$ of the $K_1(1400)$ decays 
is dominated by the $K^*\pi$. Therefore, from Fig.~\ref{gakk} 
the $\eta_1(1855)\to \bar{K}K^* \pi$ amplitude is written as
\begin{align}
\mathcal{M}_{3B}= g_{K_1(1400)\bar{K}} \left(-g_{\mu\nu} + 
\frac{p_{\mu}p_{\nu}}{M^2_{K_1}}\right) \frac{1}{p^2- M^2_{K_1} 
+ i\,M_{K_1} \Gamma_{K_1}}\,g_{K^*\pi}\,
\varepsilon^{\mu}_{\eta_1}\varepsilon^{\nu}_{K^*} \, ,
\end{align}
where $g_{K_1(1400)\bar{K}}$ is the coupling of the 
pole associated with the $\eta_1$ state to the $K_1(1400)\bar{K}$ 
channel, $g_{K^*\pi}$ is the $K_1(1400) K^*\pi$ coupling 
extracted from the $K_1(1400)\to K^*\pi$ reaction in the Review of Particle Physics (RPP)~\cite{ParticleDataGroup:2022pth},
and $\varepsilon^{\mu}_{\eta_1}$ and $\varepsilon^{\nu}_{K^*}$ 
are the polarization vectors of the $\eta_1$ and $K^*$ mesons, 
respectively.  

The differential decay width for the $\eta_1 \to \bar{K}K^*\pi$ 
process is given by 
\begin{align}
\frac{d\Gamma}{dM_{K_1\bar{K}}} = \frac{1}{(2\pi)^3}\,
\frac{p_K\,\tilde{p_{\pi}}}{4M^2_{\eta_1}}\,\vert \mathcal{M}_{3B}\vert^2\,\frac{1}{2J+1} \, ,
\label{3b}
\end{align}
where 
\begin{equation}
\tilde{p}_{\pi} = \frac{1}{2M_{K_1}}\,\lambda^{1/2}\left( 
M^2_{K_1}, m^2_{K^*}, m^2_{\pi}\right) ,
\end{equation}
and
\begin{equation}
p_K = \frac{1}{2M_{\eta_1}}\,\lambda^{1/2}\left( 
M^2_{\eta_1}, m^2_{K}, M^2_{K_1 }\right) ,
\end{equation}
with $M_{K_1}$, $m_{K^*}$, $m_{\pi}$ being the masses of the 
$K_1(1400)$, $K^*$ and $\pi$ mesons.

From Eq.~\eqref{3b} we obtain the following results for 
the $\eta_1\to \bar{K}K^*\pi$ decay width
\begin{eqnarray}
  \Gamma_{3B}&=&\left(81^{+11}_{-24}\,{\rm{MeV}}\right)^A, \notag\\
  \Gamma_{3B}&=&\left(74^{+12}_{-24}\,{\rm{MeV}}\right)^B, 
\label{3bresults}
\end{eqnarray}
where the uncertainties come from the subtraction constant 
(cutoff) used to regularize the loops in Eq.~\eqref{loopdr} 
(Eq.~\eqref{loopcut}). As can be seen from Eq.~\eqref{3bresults}, we obtain 
similar results whether we use the sets $A$ or $B$. For 
the sake of comparison to other works, we evaluate the ratio 
$\Gamma_{2B}/\Gamma_{3B}$, and get
\begin{align}
\frac{\Gamma_{2B}}{\Gamma_{3B}}= \left(0.23^{-0.08}_{+0.16}\right)^A~ \text{or} ~ \left(0.10^{-0.03}_{+0.08}\right)^B ,
\label{2b3b}
\end{align}
which is consistent to the results in Ref.~\cite{Dong:2022cuw}, 
where the $\eta_1$ is also assumed to be a $K_1(1400)\bar{K}$ 
molecular state. On the other hand, adopting the same multiquark 
configuration than the present work and Ref.~\cite{Dong:2022cuw}, 
the authors of Ref.~\cite{Yang:2022lwq} have found a 
different result for the ratio, 
$\Gamma_{2B}/\Gamma_{3B}\approx 0.03$. Nevertheless, in all 
the cases the results point out that the $\bar{K}K^*\pi$ 
three-body channel is more likely than the $\eta\eta^{\prime}$ one.

\section{The $\pi_1(1400/1600)$ dynamical generation}
\label{sec:pis}

The WT amplitudes for the pseudoscalar-axial 
vector meson interactions with $I=1$ are given by Eq.~\eqref{wt}, 
with the corresponding $C_{ij}$ coefficients listed in Table~\ref{cij1}. 
In this case, from Eq.~\eqref{bs}, we get two $\pi_1$ poles shown in 
Table~\ref{pi1600}.

{\squeezetable \begin{table}[h!]
    \caption{Poles and their corresponding couplings to the channels 
    contributing to the $PA$ interaction with $J^{PC}=1^{-+}$ and 
    $I=1$. The errors of the poles are from varying the subtraction constant within $\alpha(\mu=1\,\text{GeV})=-1.35\pm0.17$, and only the central values of the couplings are given. 
    The last row of each panel gives the central values of the peak mass and width $(M_\text{peak},\Gamma_\text{peak})$ for the corresponding pole after considering the axial-vector meson widths.
    }
    \centering
    \resizebox{\textwidth}{!}{
    \begin{tabular}{c c c c c c c}
    \hline\hline
    {Poles} (Set A) & & & ~~~~~~~~ {Channels} & & &\\
    \hline\hline
    $\bm{1.47\pm0.01- i (0.12\pm0.02)}$ \quad& $b_1\pi$ ~& $f_1(1285)\pi$ & $f_1(1420)\pi$ & $K_1(1270)\bar{K}$ ~& $a_1\eta$ ~& $K_1(1400)\bar{K}$\\ 
    $\left(--++++\right)$ & & & & & & \\
    $g_l$ & $\bm{5.22+i4.40}$ & $0.02-i 0.09$ & $0.03-i 0.05$ & $1.25+i1.27$ & $0.02-i0.12$ & $1.33+i1.63$\\
    $\bm{(1.56,0.46)}$ &  &  &  &  &  & \\
    \hline\hline
    $\bm{1.75\pm0.02- i(0.02\pm0.01)}$ \quad& $b_1\pi$ ~& $f_1(1285)\pi$ & $f_1(1420)\pi$ & $K_1(1270)\bar{K}$ ~& $a_1\eta$ ~& $K_1(1400)\bar{K}$\\ 
    $\left(---+++\right)$ & & & & & & \\
    $g_l$ & $0.10+i0.95$ & $2.73-i0.02$ & $1.89$ & $\bm{5.84-i1.85}$ & ${3.49-i0.03}$ & ${2.65-i0.53}$\\
    $\bm{(1.74, 0.30)}$ &  &  &  &  &  & \\
    \hline\hline
    {Poles} (Set B) & & & ~~~~~~~~ {Channels} & & &\\
    \hline\hline
    $\bm{1.47\pm0.01- i (0.12\pm 0.02)}$ \quad& $b_1\pi$ ~& $f_1(1285)\pi$ & $f_1(1420)\pi$ & $K_1(1270)\bar{K}$ ~& $a_1\eta$ ~& $K_1(1400)\bar{K}$\\ 
    $\left(--++++\right)$ & & & & & & \\
    $g_l$ & $\bm{5.27+i4.31}$ & $0.01-i0.03$ & $0.03-i0.06$ & $1.97-i1.81$ & $0.02-i0.08$ & $0.91+i1.07$\\
    $\bm{(1.57, 0.50)}$ &  &  &  &  &  & \\
    \hline\hline
    $\bm{1.77\pm 0.01- i(0.01\pm0.01)}$ \quad& $b_1\pi$ ~& $f_1(1285)\pi$ & $f_1(1420)\pi$ & $K_1(1270)\bar{K}$ ~& $a_1\eta$ ~& $K_1(1400)\bar{K}$\\ 
    $\left(---+++\right)$ & & & & & & \\
    $g_l$ & $0.13+i1.44$ & $1.37-i0.25$ & $2.86-i0.50$ & $\bm{4.80-i2.29}$ & ${3.53-i0.64}$ & ${4.54-i1.77}$ \\
    $\bm{(1.72, 0.20)}$ &  &  &  &  &  & \\
    \hline\hline
    \end{tabular}
    }
    \label{pi1600}
    \end{table}}

Similar to the previous section, we also provide the couplings of 
these dynamically generated states to the channels listed in Table~\ref{tab2}. 
Table~\ref{pi1600} shows a broad $\pi_1$ pole at $1.47$ GeV, and a width 
of about $0.12$ GeV.\footnote{As discussed in Section~\ref{sec:widtheffects}, the widths of the dynamically generated poles will be significantly increased once the width effects of the axial-vector mesons are taken into account; see also the discussions below.} This state is above the $b_1\pi$ and $f_1(1285)\pi$ 
thresholds. Its large width stems from the large 
coupling to the $b_1\pi$ and the fact that this channel is open for 
decaying. The $f_1(1285)\pi$ channel is also open. However, according 
to Table~\ref{cij1}, the corresponding WT term in Eq.~\eqref{wt} 
is zero for the diagonal $f_1(1285)\pi$ transition. On the other hand, 
the next $\pi_1$ pole in Table~\ref{pi1600} has a sizeable dependence on the mixing angles. 
Using set $A$, we find that pole at ${1.75}$~GeV. It couples most strongly to the $K_1(1270)\bar K$ channel, which is closed 
for decaying. Nonetheless, the state can decay into $b_1\pi$ and $f_1(1285)\pi$, 
albeit their corresponding couplings are small compared to the $K_1(1270)\bar K$ one, but still large enough to provide a sizeable width for 
the pole. In contrast, when set $B$ is adopted, the 
higher $\pi_1$ pole is now located at ${1.77}$ GeV, above the $f_1(1420) \pi$ threshold, which 
is now open. 
One might think that the width should increase since now three 
channels are open for decaying. However, although the coupling to the 
$f_1(1420)\pi$ has increased in this case, at the same time the couplings to the other open channels 
have decreased. Hence, the overall effect leads to a smaller width compared to the previous case.

\begin{figure}[h!]
\includegraphics[width=0.35\textwidth]{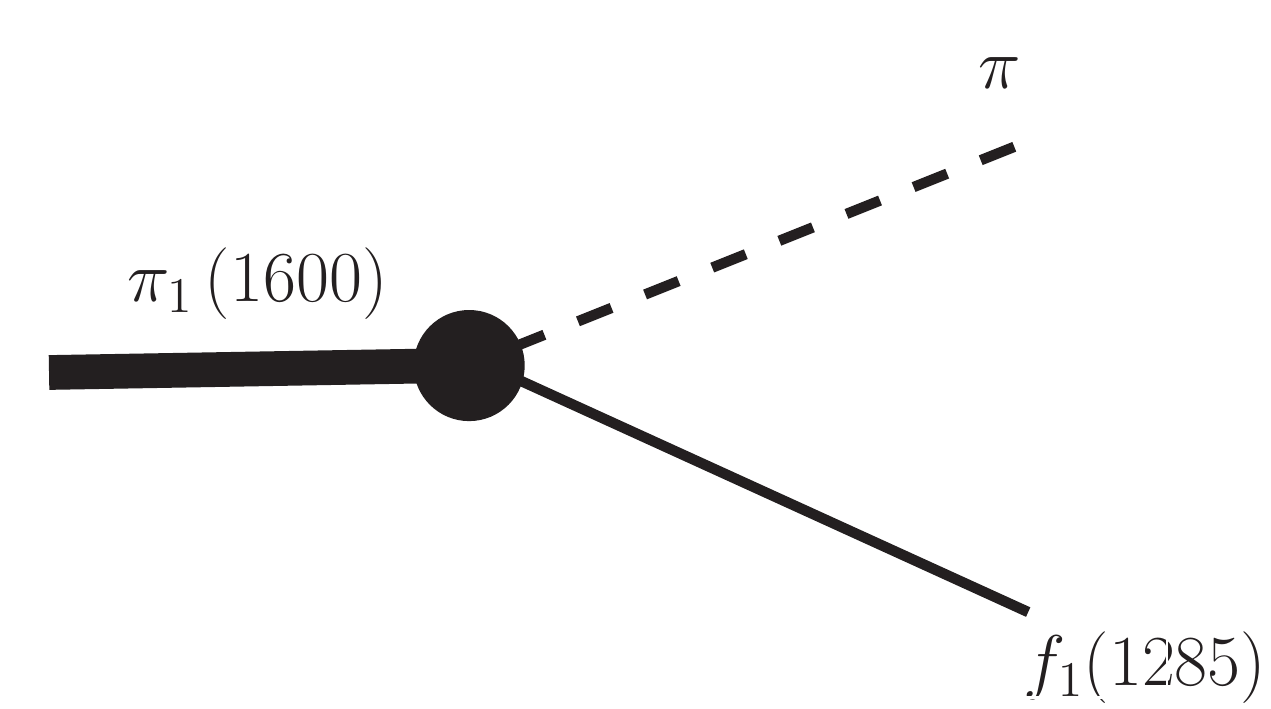}
\includegraphics[width=0.45\textwidth]{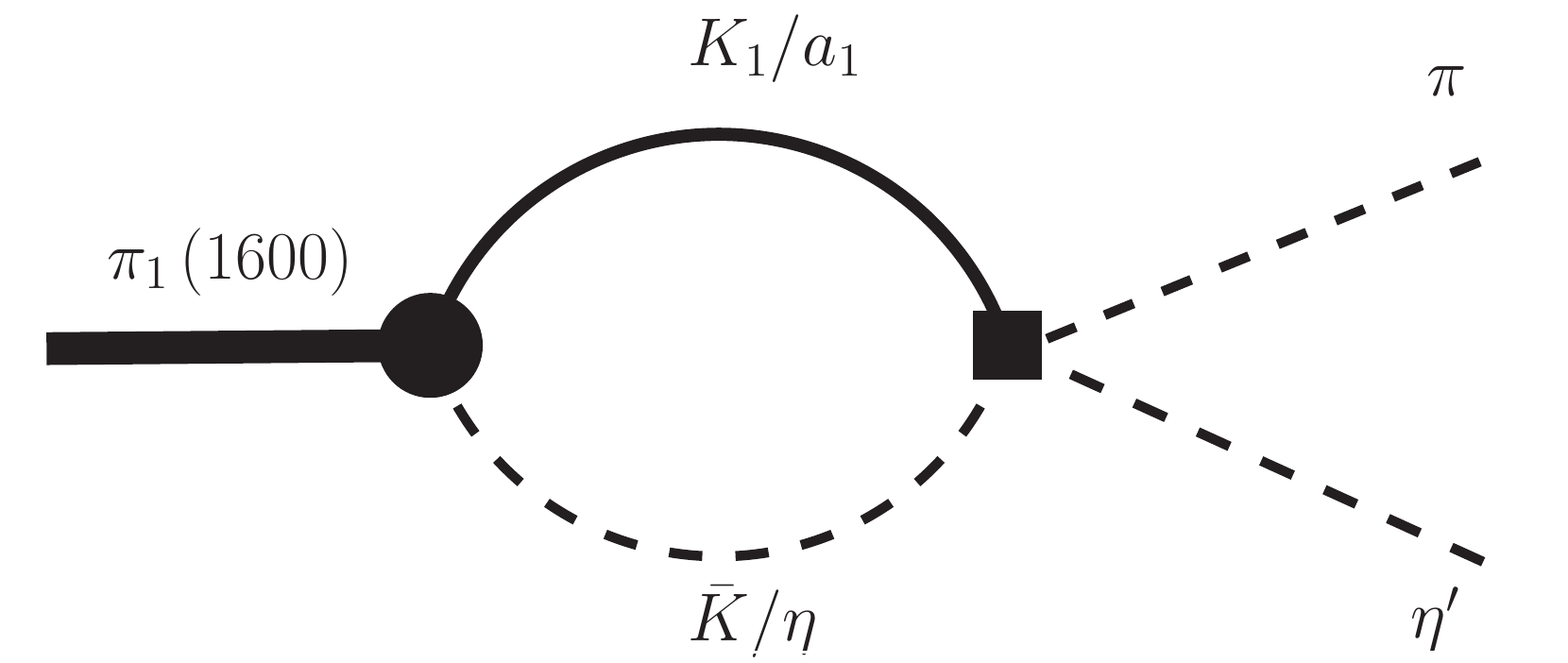}
\caption{a) Diagram corresponding to the $\pi_1(1600) \to f_1(1285)\pi$ 
reaction, and b) the $\pi_1(1600) \to 
\eta^{\prime}\pi$ decay also via the ${AP}$ loop. The filled circles represent the effective couplings of the $\pi_1$ to the $AP$ meson pairs calculated from the residues.
{The rectangles}
are the $AP\to \eta'\pi$ transition amplitudes at tree level. } \label{gapi600}
\end{figure}

The lower pole mass is slightly higher than the mass of the $\pi_1(1400)$ state listed 
in RPP, $(1354\pm25)$~MeV~\cite{ParticleDataGroup:2022pth}. 
Notice that we use the same subtraction constant for all channels. In principle, it can take different values and lead to a shift of the poles.
In addition, we did not include in the loops 
the $b_1$ width, that is relatively large and whose 
effects could influence the pole position. However, it is expected to
affect more the imaginary part of the pole than the real 
one (see Fig.~\ref{fig:isovector.1} below). 
We can get a rough estimate of this change by adding 
the $b_1$ width to the previous result for $\textrm{Im}(z_1)$, 
with $z_1$ the lower $\pi_1$ pole, {\it i.e.},
\begin{equation}
\Gamma_{b_1}+ 2\textrm{Im}(z_1) \approx 0.4\,\, \textrm{GeV}\, ,\label{eq:pi1Wid}
\end{equation}
which is close to the $\pi_1(1400)$ width reported in RPP, $(330 \pm 35)$~MeV~\cite{ParticleDataGroup:2022pth}. 
From these results, we are led to claim that the lower $\pi_1$ pole may 
explain the $\pi_1(1400)$ resonance; in other words, the 
$\pi_1(1400)$ is suitably described in our approach as a dynamically 
generated state with the $b_1 \pi$ as its main component.

Alternatively, following the prescription used in Section \ref{sec:eta}, we can also study the changes in the results caused by the inclusion of the finite widths for the axial-vector mesons by looking at the line shape for the relevant $T$-matrix elements. In Fig.~\ref{fig:isovector.1} we show the line shapes for the $T$-matrix element corresponding to the elastic $b_1\pi\to b_1\pi$ transition, which is the one we would expect the lower pole in Table~\ref{pi1600} manifests most due to its large coupling to the $b_1\pi$ channel. 
It becomes clear that the bumps become broader when the widths of axial-vector mesons are taken into account.
A similar behavior can be seen in Fig.~\ref{fig:isovector.2} for the $T$-matrix element associated with the scattering of $K_1\left(1270\right)\bar{K}$, which is the channel to which the higher $\pi_1$ pole couples most strongly.
The peak mass and width extracted from the line shape of the diagonal $T$-matrix element for the dominantly coupled channel are also listed in Table~\ref{pi1600} when the axial-vector meson widths are considered.

\begin{figure}[htb]
  \centering  
  \subfigure[Modulus square of elastic
  $b_1\pi$ scattering
]{
  \label{fig:isovector.1}
  \includegraphics[width=0.45\textwidth]{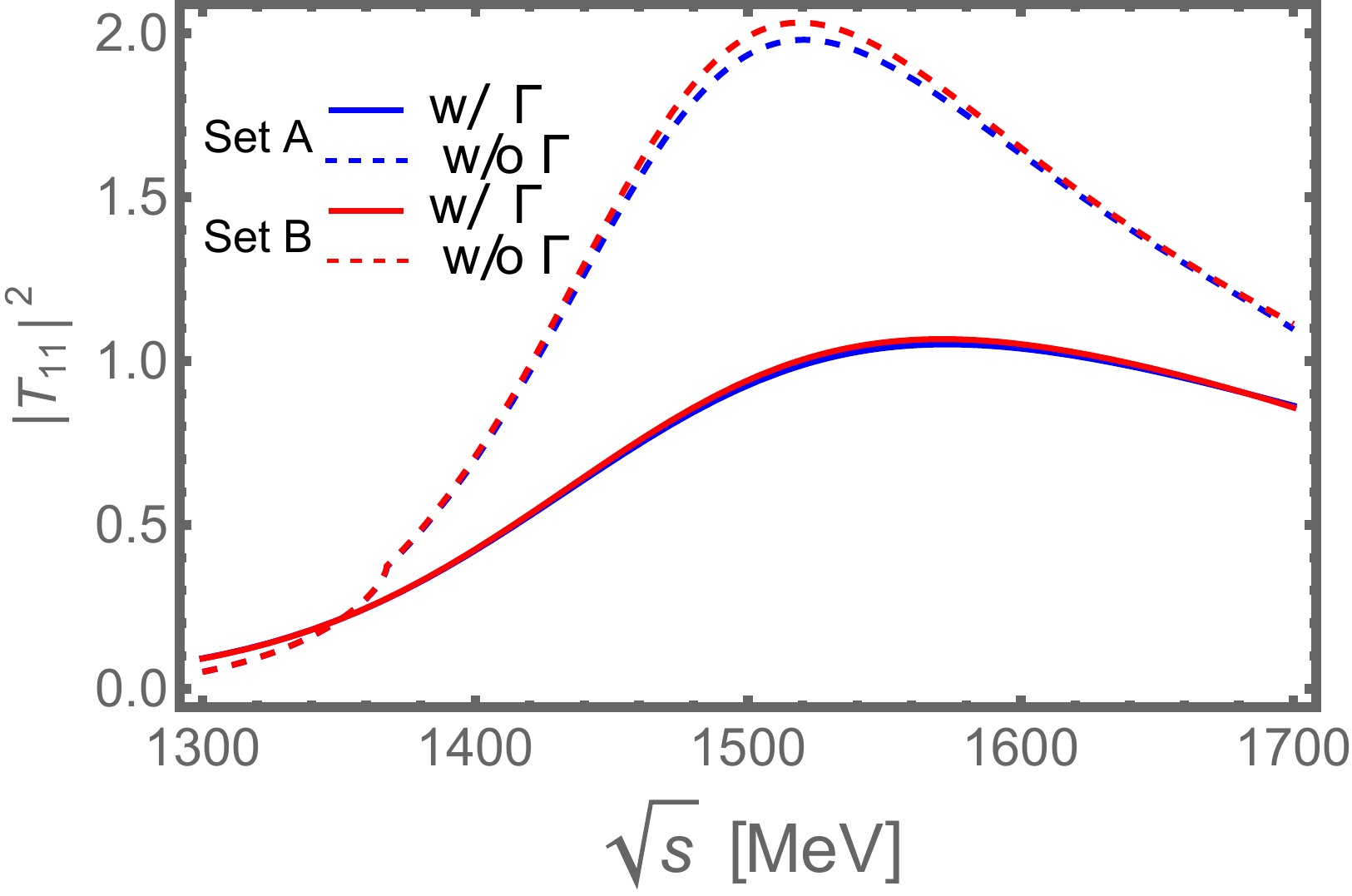}}
  \subfigure[Modulus square of elastic $K_1\left(1270 \right)\bar{K}$ scattering]{
  \label{fig:isovector.2}
  \includegraphics[width=0.43\textwidth]{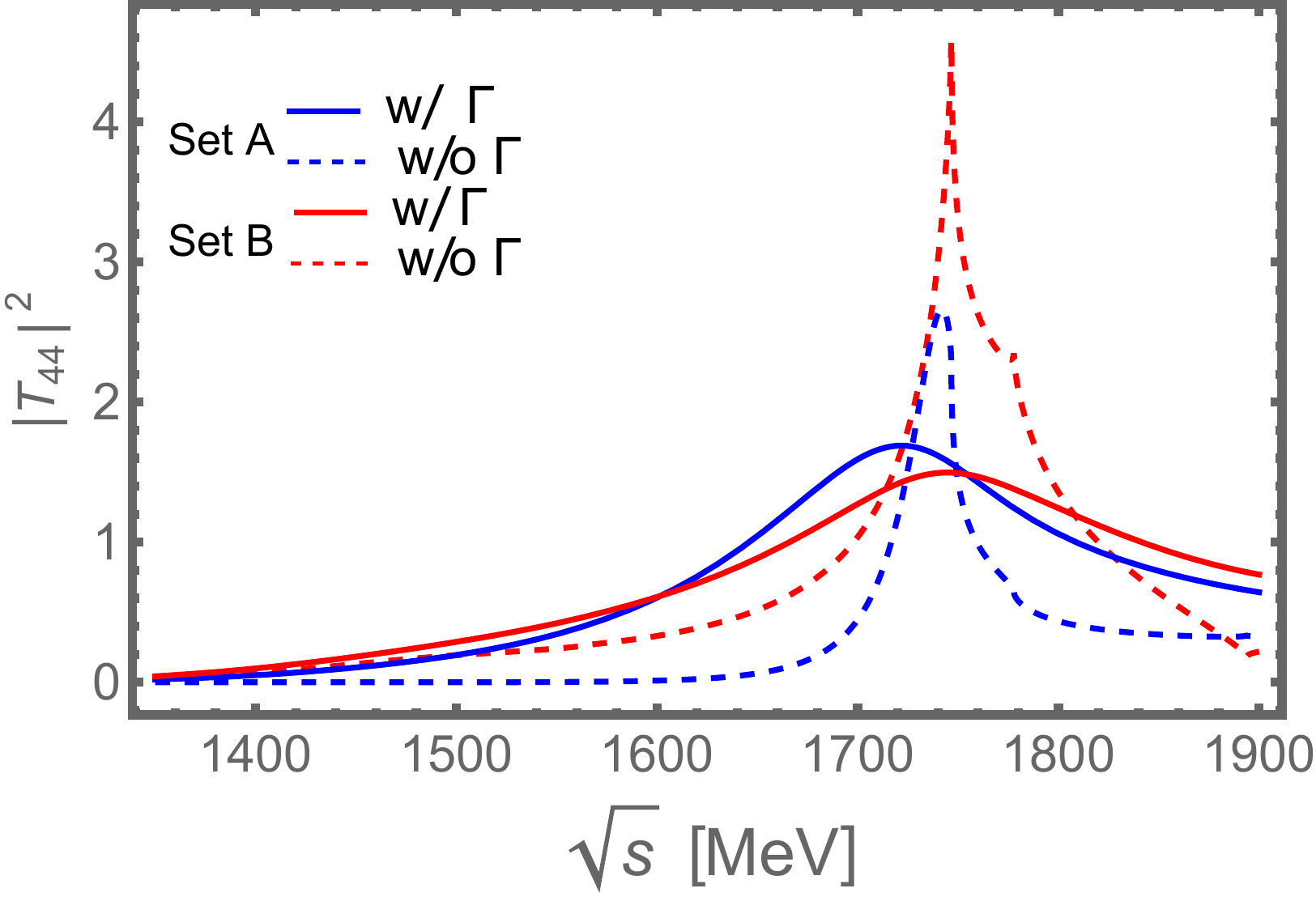}}
 \caption{The dashed and solid lines correspond to zero and full widths of the axial-vector mesons in $G^{Cut}$. }
  \label{fig:isovector}
\end{figure}

The higher $\pi_1$ pole, denoted now by $z_2$, has a mass consistent with that of the $\pi_1(1600)$, whose pole mass has been reported to be $\left(1623 \pm 47_{-75}^{+24}\right)$~MeV in Ref.~\cite{Kopf:2020yoa} and $(1564 \pm 24 \pm 86)$~MeV in Ref.~\cite{JPAC:2018zyd}. It can decay into the $\eta^{\prime}\pi$ and $f_1(1285)\pi$ 
channels. The corresponding  diagrams for both amplitudes are illustrated in Fig.~\ref{gapi600}, 
from which we have 
\begin{equation}
\mathcal{M}_{f_1(1285)\pi} = g_{f_1\left(1285 \right)\pi} \varepsilon_{\eta_1}\cdot \varepsilon_{f_1}\,, 
\end{equation} 
and 
\begin{equation}
\mathcal{M}_{\eta^{\prime}\pi} = g_{K_1\bar{K}}G_{K_1\bar{K}}V_{K_1\bar{K},\eta^{\prime}\pi} \cdot \varepsilon_{\eta_1}+g_{a_1\eta}G_{a_1\eta}V_{a_1\eta,\eta^{\prime}\pi}\cdot\varepsilon_{\eta_1}\, ,
\end{equation} 
with $\varepsilon_{\eta_1}$ and $\varepsilon_{f_1}$ the polarization vectors of the $\eta_1$ and $f_1\left( 1285\right)$ mesons.
Here $g_{f_1\left(1285 \right)\pi}$, $g_{K_1\bar{K}}$ and  $g_{a_1\eta}$ are the effective coupling of the $z_2$ pole to the corresponding couplings, and $G_{K_1\bar{K}}$ and $G_{a_1\eta}$
are the loops involving the $K_1\bar{K}$ and $a_1\eta$ mesons, respectively.
Notice that the effective couplings are computed from the residues of the $T$ matrix elements; thus they contain contributions from all coupled channels.

In order to compare our findings with the experimental information, we evaluate 
the ratio
\begin{equation}
\mathcal{R}_1= \frac{|\mathcal{M}_{f_1(1285)\pi}|^2\,q}
{|\mathcal{M}_{\eta^{\prime}\pi}|^2\,\tilde{q}}\, ,
\label{ratio1}
\end{equation}
where $q$ and $\tilde{q}$ are the momentum in the c.m. frame of the $f_1(1285)\pi$ and $\eta^{\prime}\pi$ pairs, 
respectively. Numerically, Eq.~\eqref{ratio1} gives
  \begin{align}
  \mathcal{R}_{1}=\left\{\begin{array}{l}
    \left(2.4^{+0.8}_{-0.6}\right)^A,\\
    \left(2.1^{+0.4}_{-0.3}\right)^B.
  \end{array}\right.
  \label{ratios}
  \end{align}
The ratio is slightly bigger 
for the mixing angles in the set $A$. Nevertheless, the result in 
Eq.~\eqref{ratios} is consistent to the corresponding ratio  $3.80\pm 0.78$ reported by the E852 Collaboration~\cite{E852:2004gpn}. This good agreement with the 
experimental data supports the molecular picture for 
the $\pi_1(1600)$ state.

\section{Dynamical generation in $I=1/2$ sector}
\label{sec:ihalf}

In the $I=1/2$ sector, the corresponding WT amplitudes are given 
by Eq.~\eqref{wt} with the $C_{ij}$ coefficients 
given in Tables~\ref{cijha} and \ref{cijhb}. For 
each case, we have found two poles for parameter sets $A$ and $B$, 
as shown in Table \ref{polesiha} and \ref{polesihb}. 
{\squeezetable \begin{table}[h!]
      \caption{Poles and their corresponding couplings to the channels 
      contributing to the $PA$ interaction with $J^P=1^-$. Here the flavor-neutral axial mesons have $J^{PC}=1^{++}$. The errors of the poles are from varying the subtraction constant within $\alpha(\mu=1\,\text{GeV})=-1.35\pm0.17$, and only the central values of the couplings are given. The last row of each panel gives the central values of the peak mass and width $(M_\text{peak},\Gamma_\text{peak})$ for the corresponding pole after considering the axial-vector meson widths.} 
      \centering
      \begin{tabular}{ c c c c c c}
      \hline\hline
      {Poles} (Set A) & & ~~~~~~~~~{Channels} & &\\
      \hline\hline
      $\bm{1.69\pm 0.02}$ ~& $a_1 K$~& $f_1(1285)K$~& $K_1(1270)\eta$~& $f_1(1420)K$~&~ $K_1(1400)\eta$\\
      $(+++++)$~& ~& ~& ~& ~~&~ \\
      $g_l$ ~& $\bm{6.89}$ & $0.89$ & $3.75$ & $0.54$ & $2.10$\\
      $\bm{(1.70, 0.28)}$ ~& &  & &  & \\
      \hline\hline
      {Poles} (Set B) & & ~~~~~~~~~{Channels} & &\\
      \hline\hline
      $\bm{1.70\pm 0.02}$ ~& $a_1 K$~& $f_1(1285)K$~& $K_1(1270)\eta$~& $f_1(1420)K$~&~ $K_1(1400)\eta$\\
      $(+++++)$~& ~& ~& ~& ~~&~ \\
      $g_l$ ~& $\bm{6.58}$ & $0.25$ & $2.45$ & $0.27$ & $3.15$\\
      $\bm{(1.70, 0.30)}$ ~& &  & &  & \\
      \hline\hline
      \end{tabular}
      \label{polesiha}
      \end{table}}

{\squeezetable \begin{table}[h!]
      \caption{Poles and their corresponding couplings to the channels 
      contributing to the $PA$ interaction with $J^P=1^-$. Here the flavor-neutral axial mesons have $J^{PC}=1^{+-}$. The errors of the poles are from varying the subtraction constant within $\alpha(\mu=1\,\text{GeV})=-1.35\pm0.17$, and only the central values of the couplings are given. The last row of each panel gives the central values of the peak mass and width $(M_\text{peak},\Gamma_\text{peak})$ for the corresponding pole after considering the axial-vector meson widths.}
      \centering
      \begin{tabular}{cccccc}
      \hline\hline
      {Poles} (Set A) & & {Channels} & &\\
      \hline\hline
      $\bm{1.70\pm0.02}$ & $h_1(1170) K$ & $b_1 K$ & $K_1(1270)\eta$ & $h_1(1415) K$ &  $K_1(1400)\eta$\\
      $(-++++)$~& ~& ~& ~& ~&~ \\
      $g_l$~& $0.20$ & $\bm{6.46}$ & $2.38-i0.01$ & $0.50$ & $3.21-i0.02$ \\
      $\bm{(1.70, 0.14)}$ ~& &  & &  & \\
      \hline\hline
      {Poles} (Set B) & & {Channels} & &\\
      \hline\hline
      $\bm{1.69\pm 0.02}$ ~& $h_1(1170) K$ & $b_1 K$ & $K_1(1270)\eta$ & $h_1(1415) K$ &  $K_1(1400)\eta$\\
      $(-++++)$~& ~& ~& ~& ~~&~ \\
      $g_l$~& $0.55-i0.01$ & $\bm{6.78+i0.02}$ & $3.69-i0.06$ & $0.83-i0.01$ & $2.17-i0.04$ \\
      $\bm{(1.70, 0.14)}$ ~& &  & &  & \\
      \hline\hline
      \end{tabular}
      \label{polesihb}
\end{table}}

Similarly to the previous cases, the poles are located on the same Riemann sheets in both sets of mixing angles. The interactions in the $a_1 K$ and $b_1 K$ channels are strong to generate a bound state in each of them. The existence of a lower $h_1\left( 1170\right)K$ channel below the $b_1K$ threshold moves the pole in Table~\ref{polesihb} to Riemann sheet $\left( -++++\right)$. It has a nonzero imaginary part of a few MeV, which is not shown in the table due to precision.

As discussed before, the $I=1/2$ poles in Tables ~\ref{polesiha} and 
\ref{polesihb} will receive sizeable widths once the width effects of the axial-vector mesons are taken into account, and it is expected that the widths are of the order of a few hundred MeV, like those of the $b_1$ and $a_1$ mesons.
The peak mass and width for each pole are also listed in the tables when the axial-vector meson widths are considered.
Although we neglected the transitions between the $A_1P$ and $B_1P$ sectors as discussed around Eq.~\eqref{eq:a1b1} in Section~\ref{sec:formalism}, strange mesons are not $C$-parity eigenstates and the two dynamically generated $I=1/2$ $1^-$ states will inevitably mix. The two mixed states together could correspond to the $1^-$ $K^{\ast}\left(1680\right)$ structure~\cite{ParticleDataGroup:2022pth}.
%

\section{Conclusions}
\label{sec:conc}

We have studied the interactions between the pseudoscalar 
and axial-vector mesons in coupled channels with $J^{PC}=1^{-(+)}$ 
quantum numbers for the isospin $0$, $1$, and $1/2$ sectors. 
Using the chiral unitary approach, we describe the interaction 
with the Weinberg-Tomozawa term derived from chiral Lagrangians. 
The transition amplitudes among all the 
relevant channels are unitarized using the Bethe-Salpeter 
equation from which resonances (bound states) manifest themselves 
as poles on the (un)physical Riemann sheets of the complex energy 
plane. 

We consider the physical isoscalar axial-vector states as mixtures of the corresponding SU(3) singlets and octets. 
In addition, the $K_1(1270)$ and $K_1(1400)$ physical states 
are also mixtures of the $K_{1A}$ and $K_{1B}$ mesons, which 
are the strange partners of the $a_1$ and $b_1$ resonances, respectively. 
We group into two sets, called $A$ and $B$, the mixing 
angles accounting for such mechanisms and 
investigate their influence on the pole positions.

According to our findings, we obtain poles with $J^{P(C)}=1^{-(+)}$  quantum numbers in the energy range from $1.30$ 
to $2.00$ GeV, in each isospin sector studied ($I=0,1,1/2$). The $1^{-+}$ quantum numbers are exotic in the sense that they cannot be formed from a pair of quark and antiquark. In particular, 
we have found an isoscalar state that may correspond to 
the $\eta_1(1855)$ state, newly observed by the BESIII Collaboration~\cite{BESIII:2022riz}. 
In addition, we have also found two dynamically generated 
isovector states that we assign to be the $\pi_1(1400)$ and $\pi_1(1600)$ 
resonances. Hence, within our formalism, they are dynamically 
generated through the pseudoscalar-axial vector meson 
interactions, with the $\eta_1(1855)$ state coupling mostly to 
$K_1(1400)\bar{K}$ channel, while the $\pi_1(1400)$ couples strongly 
to the $b_1\pi$, and $\pi_1(1600)$ structure couples  most strongly to the 
$K_1(1270)\bar{K}$. 
We also find two $I=1/2$ $J^P=1^-$ states with a mass around 1.7~GeV. They combined together could be responsible to the observed $K^*(1680)$ structure.

In addition, we also evaluate the decays of the $\eta_1(1855)$ and the $\pi_1(1600)$. We find that the three-body decay channel $\bar K K^*\pi$ has a significantly larger branching fraction than the $\eta'\eta$, which is the channel where the observation of the $\eta_1(1855)$ was made. The obtained ratio between the 
$\pi_1(1600)\to f_1(1285)\pi$ and $\pi_1(1600)\to \eta^{\prime}\pi$ 
decays, given by 
Eq.~\eqref{ratios}, is consistent with the corresponding 
experimental value. 

We suggest searching for two additional $\eta_1$ exotic mesons with masses of about 1.4 and 1.7~GeV, respectively. In particular, the latter should be relatively narrow with a width around 0.1~GeV and one of its main decay channels is $K\bar K \pi\pi$.

\begin{acknowledgments}
M.~J.~Y is grateful to Shuang-Shi Fang and M.~P.~Valderrama for valuable discussions. 
This project is supported in part by the National Natural Science Foundation of China (NSFC) under Grants No.~12125507, No. 11835015, and No.~12047503; by the China Postdoctoral Science Foundation under Grant No. 2022M713229; by the NSFC and the Deutsche Forschungsgemeinschaft (DFG) through the funds provided to the Sino-German Collaborative Research Center TRR110 “Symmetries and the Emergence of Structure in QCD” (NSFC Grant No.~12070131001, DFG Project-ID~196253076); and by the Chinese Academy of Sciences under Grant No.~XDB34030000.
\end{acknowledgments}

\bibliography{eta1refs.bib}

\end{document}